\documentclass{article}

\usepackage{arxiv}

\usepackage{xcolor}

\usepackage[utf8]{inputenc} 
\usepackage[T1]{fontenc}    
\usepackage{hyperref}       
\usepackage{url}            
\usepackage{booktabs}       
\usepackage{amsfonts}       
\usepackage{nicefrac}       
\usepackage{microtype}      
\usepackage{lipsum}         
\usepackage{graphicx}
\usepackage{natbib}
\usepackage{doi}
\usepackage{hyperref}
\usepackage{float}
\usepackage{url}            
\usepackage{booktabs}       
\usepackage{amsfonts}       
\usepackage{amssymb}        
\usepackage{nicefrac}       
\usepackage{microtype}      
\usepackage{bm}				
\usepackage{mathrsfs}		
\usepackage{cite}			
\usepackage{graphicx}		
\usepackage{mathtools}		
\usepackage{algorithmic}	
\usepackage{algorithm}
\usepackage{amsthm}			
\usepackage{enumitem}		
\usepackage{multirow}       
\usepackage{tabularx}	    
\usepackage{hhline}
\usepackage{arydshln}		
\usepackage{wrapfig}
\usepackage{lipsum}
\usepackage{diagbox}
\usepackage{soul}
\usepackage{amsthm}
\usepackage{enumitem}
\theoremstyle{plain} 
\newtheorem{proposition}{Proposition}

\newtheorem{lemma}{Lemma}

\newtheorem{assumption}{Assumption}

\title{Robust and Fast Bass Local Volatility}

\author{
    {\large Hao Qin$^{\dagger}$, \href{https://orcid.org/0009-0009-4556-8664}{\includegraphics[scale=0.06]{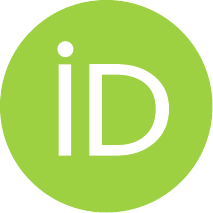}\hspace{1mm}Charlie Che$^{\dagger\ddagger}$}, Ruozhong Yang$^{\dagger}$, and Liming Feng$^{\dagger}$\thanks{Corresponding author. Email: fenglm@illinois.edu.}}\\[1ex]
    $^{\dagger}$Department of Industrial and Enterprise Systems Engineering, University of Illinois Urbana-Champaign, Illinois, United States\\
    $^{\ddagger}$Quantitative Trading \& Research, JPMorgan Chase \& Co., New York, United States
}
\date{} 

\renewcommand{\shorttitle}{Robust and Fast Bass Local Volatility}

\begin{document}
\maketitle

\begin{abstract}
The Bass Local Volatility Model, as studied in \citep{henry2021bass}, stands out for its ability to eliminate the need for interpolation between maturities. This offers a significant advantage over traditional local volatility models. However, its performance highly depends on accurate construction of risk neutral densities and the corresponding marginal distributions and efficient numerical convolutions which are necessary when solving the associated fixed point problems. In this paper, we propose a new approach combining local quadratic estimation and lognormal mixture tails for the construction of risk neutral densities. We investigate computational efficiency of trapezoidal rule based schemes for numerical convolutions and show that they outperform commonly used Gauss-Hermite quadrature. We demonstrate the performance of the proposed method, both in standard option pricing models, as well as through a detailed market case study.

\end{abstract}
\vspace{1em} 

\vspace{1em} 

\noindent\textbf{Key Messages:}
\begin{itemize}
\item Development of high quality arbitrage-free risk neutral densities necessary for Bass local volatility implementation
\item Theoretical and numerical evidence of the trapezoidal rule’s superiority over Gauss-Hermite quadrature for numerical convolution in the Bass local volatility implementation
\end{itemize}

\section{Introduction}

In derivatives pricing, local volatility (LV) models have been widely adopted, particularly in applications involving the valuation of exotic options \citep{dupire.1994, derman1996local, coleman2001reconstructing, bouzoubaa2010exotic, kotze2015implied}. Dupire's formulation of LV models provides a deterministic framework where instantaneous volatility is a function of both the asset price and time. Since Dupire's seminal work, this framework has become a key tool and industry standard for characterizing the dynamics of underlying asset prices. However, practical application of such models faces challenges due to the lack of observable vanilla option prices across all strikes and maturities. One of the main challenges is the need for an arbitrage-free interpolation scheme for volatilities at unobserved maturities. The time interpolation and related extrapolation can introduce instabilities and make the model highly sensitive to variations in market data.

This difficulty can be effectively addressed by the Bass-LV construction proposed by \citet{henry2021bass}, which extends the Bass martingale of \citet{bass1983} to a multi-marginal setting and calibrates exactly to discrete marginals. This construction is distinct from, though related to, the Martingale Benamou--Brenier framework of \citet{backhoff2017martingale}, which generalizes the Bass martingale from an optimal transport perspective. The Bass-LV model leverages the martingale property of the asset price process to ensure the absence of calendar arbitrage, thereby circumventing the need for direct time interpolation of volatilities. This construction is particularly advantageous because it aligns with the martingale condition, which is a fundamental requirement for no-arbitrage pricing in derivatives markets.

The Bass-LV model is particularly well-suited for pricing a wide array of exotic payoffs, such as autocalls, forward-start options, lookback options, and Asian options. Its accuracy and flexibility in addressing the complex and diverse needs of exotic option pricing make Bass-LV a competitive choice in volatility modeling. The model takes options-implied risk neutral densities as inputs and calibrates the spot price process by solving a fixed-point problem numerically. The calibrated spot price process is trivial to simulate. Monte Carlo methods can then be easily used to evaluate any financial option contract.

Recent theoretical work by \citep{acciaio2023calibration} further advances the Bass-LV theory, demonstrating a linear convergence rate of the numerical scheme in the fixed-point algorithm. \citep{tschiderer2024q} extends the Bass-LV construction by replacing the Gaussian transition kernel with an arbitrary reference measure, which provides more flexibility in fitting different financial assets. \citep{joseph2023measure} introduces the measure-preserving martingale Sinkhorn (MPMS) algorithm to compute the Bass martingale and associated dual potentials for prescribed marginals; by linking the Bass martingale with semimartingale optimal transport and exploiting a PDE-based dual formulation of the Martingale Benamou-Brenier problem, the algorithm provides an alternative iterative scheme that is shown to be equivalent to the fixed-point method of \citep{henry2021bass}, with convergence guaranteed in arbitrary dimension. Moreover, \citep{backhoff2023bass} and \citep{backhoff2023structure} give general theoretical results by analyzing the dual formulation of the Bass martingale, and strengthen the model's theoretical foundation.

The concept of leveraging martingales in Bass-LV is indeed a part of Martingale Optimal Transport (MOT), which extends the classical Optimal Transport problem \citep{monge.1781} by incorporating the martingale property. Beyond the Bass-LV construction, a variety of research has explored alternative MOT frameworks to address different problem settings. For example, \citep{henry2019martingale} gives a new class of stochastic volatility models by combining martingale Schrodinger bridges with the Sinkhorn algorithm from \citep{de2019building}.  \citep{eckstein2021computation} gives a feed-forward neural network formulation of MOT problems that enables a neural network solution for problems constructed in high dimensional space with multiple assets. \citep{henry2019martingaleNN}
proposes a primal-dual algorithm for solving MOT problems that leverages the capabilities  of generative adversarial networks. Some other related works can be seen in \citep{hobson2012robust.ref}, \citep{beiglbock2013model.ref}, \citep{dolinsky2014martingale.ref}, \citep{henry2016explicit.ref}, \citep{guo2017local.ref}, \citep{henry2017model.ref}, and \citep{nutz2023martingale.ref}.

The practical significance of the Bass--LV framework demands a numerical implementation that is both robust and fast. Yet the existing literature lacks an in-depth study of the underlying numerical schemes, as well as a systematic approach to constructing well behaved risk-neutral density inputs that are compatible with these numerical schemes. 

First, the calibration relies on a fixed-point algorithm that involves multiple convolutions, which are often implemented using Gauss--Hermite quadrature. Preliminary results suggest that using a small number of quadrature points compromises accuracy, while increasing the number substantially raises the computational time. 
Most of the recent works mentioned above focus primarily on the theoretical and algorithmic study of the Bass--LV fixed-point problem or the MBB construction. Our focus, by contrast, is on an equally important but largely overlooked layer: the numerical efficiency of a practical implementation. In market applications, the dominant computational cost and one of the primary sources of instability arise from repeated evaluation of the convolution operators within the fixed-point iteration. We therefore provide an optimal parameter selection rule and convergence analysis for trapezoidal-rule based numerical convolutions under given smoothness of the fitted marginals, and demonstrate both theoretically and numerically that it achieves a superior accuracy--runtime tradeoff relative to the commonly used Gauss--Hermite quadrature in this setting.

Second, computing implied risk-neutral densities and the corresponding marginals commonly relies on interpolation/extrapolation under the Breeden--Litzenberger identity. A broad literature recovers risk-neutral densities using parametric smile specifications \citep{shimko1993bounds,gatheral2014arbitrage}, non/semi-parametric smoothing under shape restrictions \citep{ait1998nonparametric,ait2003nonparametric,benko2007extracting}, or hybrid approaches that fit a flexible middle region and attach parametric tails \citep{gemmill2000useful,brunner2003arbitrage} (see \citealp{figlewski2018risk} for a review). 

In the Bass--LV setting, the marginal densities serve as inputs to repeated numerical convolutions within the fixed-point iteration, so their construction must be designed jointly with the downstream numerical scheme. In particular, the CDF and its inverse must be globally computable and numerically stable throughout the fixed-point iteration. The regularity enforced at the interior--tail stitching points is not merely a cosmetic smoothness choice: it determines the convergence rate of the numerical convolution and therefore must be selected in conjunction with the analysis in Section~\ref{sec: optimality theory}. In this spirit, we propose a unified risk-neutral density construction pipeline that integrates existing research works into a single coherent framework with desired regularity. In the liquid interior, the density is estimated via adaptive local quadratic smoothing following \citet{benko2007extracting}. In the tails, it's completed using lognormal-mixture distributions within the framework of \citet{brunner2003arbitrage}. At the interior--tail stitching points, boundary and monotonicity constraints are enforced to an order chosen in conjunction with the requirements of the numerical convolution schemes and the convergence theory developed in Section~\ref{sec: optimality theory}, rather than being imposed ad hoc.

Beyond its modeling advantages, the Bass-LV framework possesses a distinctive practical
feature that has received little attention in the literature.
In standard implied-volatility surface fitting, when an off-grid maturity is
needed---whether for pricing an OTC European option or for generating the
local volatility function on a PDE/Monte Carlo time grid---the entire surface
must typically be re-smoothed or re-calibrated to incorporate the new slice.
By contrast, the Bass--LV calibration is performed once on the set of quoted
maturities $\{T_i\}$, and the resulting fixed-point solutions $\{F_{W_{T_i}}\}$
are stored alongside the fitted surface.  Any intermediate maturity
$t\in(T_i,T_{i+1})$ can then be recovered on demand via heat-kernel
convolution, yielding an implied-volatility slice that is guaranteed to be
free of both calendar and butterfly arbitrage.
This separation of calibration and evaluation makes the Bass--LV
construction particularly well suited to production environments where local
volatility evaluations are required at a large number of time points. In this paper, we formally establish that the Bass--LV time interpolation scheme preserves both butterfly and calendar-arbitrage freeness.

In the following, Section~\ref{sec:bass-lv} reviews the mathematical
background of the Bass--LV construction and details its role as an
arbitrage-free time interpolation scheme, contrasting it with standard
industry practice. Section \ref{sec:rnd} generates an arbitrage-free risk-neutral density from observable option prices via adaptive local quadratic estimation in the middle region and lognormal-mixture tail completion, together with additional boundary/monotonicity and calendar-consistency checks required by the Bass setting. Section \ref{sec: optimality theory} analyzes the optimal parameter selection and convergence of the trapezoidal rule for the numerical convolution operators in Bass-LV implementation and compares it to Gauss--Hermite quadrature. Section \ref{sec:lab} presents numerical experiments in standard models and a detailed market case study.

\section{Bass-LV}
\label{sec:bass-lv}
The Bass-LV model stands out as a Markov model that achieves precise calibration to the underlying asset price distributions $\mu_1, \ldots, \mu_n$ implied from market prices of European vanilla options across a  range of maturities $T_0=0 \leq T_1 < \cdots < T_n=T$. The core of the Bass-LV model lies in the extension of the Bass martingale construction within the context of the Skorokhod embedding problem. To be specific, given probability distributions $\mu_1<\cdots<\mu_{n}$ in $\mathcal{P}(\mathbb{R})$ that are ordered in convex sense, the objective is to construct a martingale $M_t$ such that $M_t= f_t(W_t,t), t \in [T_i,T_{i+1}]$, $M_{T_i} \sim \mu_i$, and $M_{T_{i+1}} \sim \mu_{i+1}$. Here, $\mathcal{P}(\mathbb{R})$ denotes the collection of all probability measures defined on the real line $\mathbb{R}$, and $W_t$ denotes a predictable RCLL process (right-continuous with left limits) such that $W_t=W_{T_i}+B_t-B_{T_i}$ for all $t \in\left[T_i, T_{i+1}\right)$ and for all $i=0, \cdots, n-1$, with $\left(B_t\right)_{0 \leq t \leq T}$ being a standard Brownian motion. Notice that the generalization of the Bass-LV from the classical Bass Martingale is indeed not trivial.  For the classical Bass Martingale in the one-dimensional case, the initial marginal distribution is a Dirac measure $\delta_m$, where $m$ is the mean of some random variable $\nu$. It turns out the solution to the martingale optimal transport problem (also known as stretched Brownian Motion) is equivalent to finding the martingale that closely tracks a baseline Brownian motion while respecting the initial and terminal marginals.  As such, finding the solution comes down to finding a monotone increasing function $f: \mathbb{R}\mapsto\mathbb{R}$ such that $f(\gamma)=\nu$ where $\gamma$ is a standard normal random variable on $\mathbb{R}$.  The martingale $M_t$ can be defined as
\begin{align}
    M_t:=\mathbb{E}[f(B_1)|\mathscr{F}]=\mathbb{E}[f(B_1)|B_t]=f_t(B_t).
\end{align}

\begin{figure}[ht]
\centering
\begin{minipage}{0.45\linewidth}
    \includegraphics[width=\linewidth]{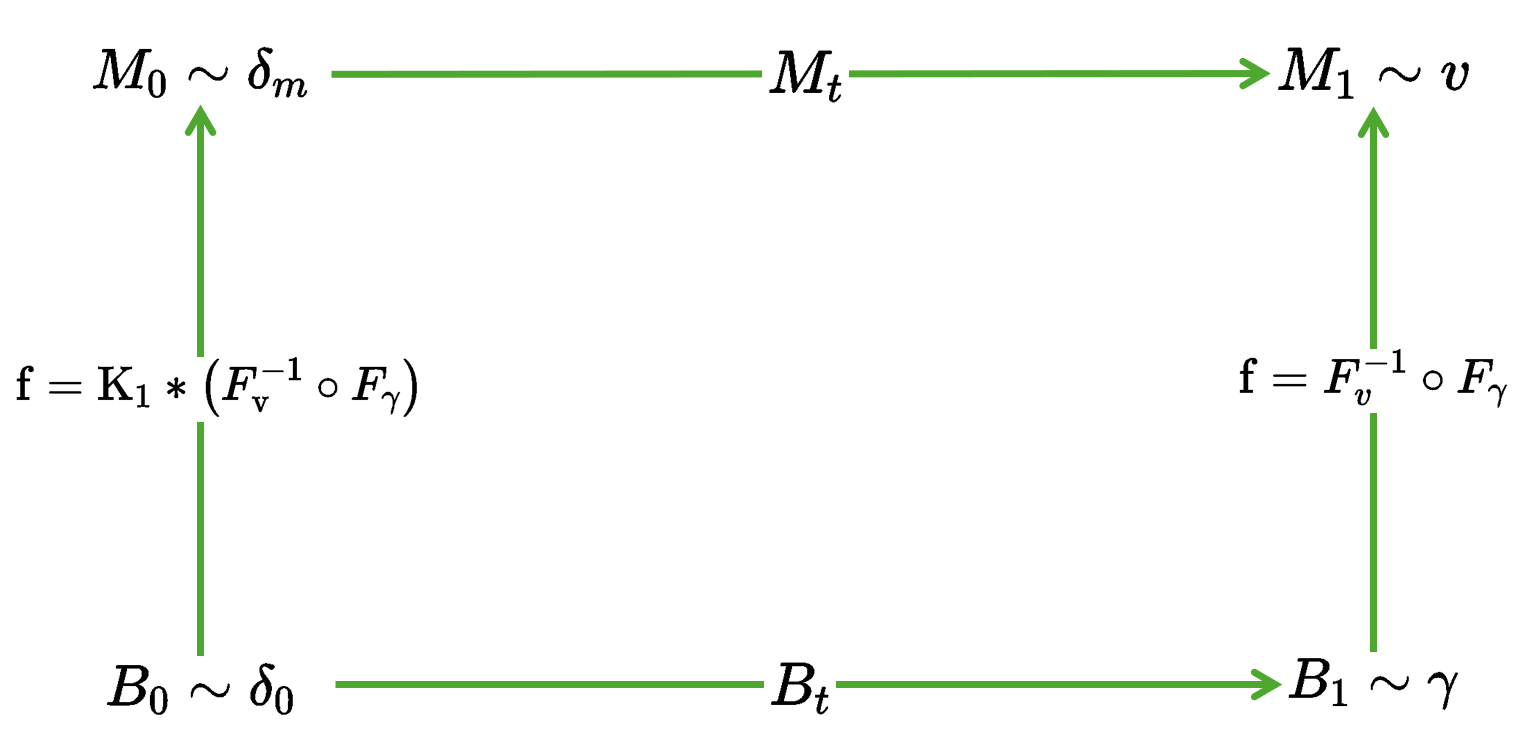}
    \caption{Bass Martingale}
    \label{fig:Bass Martingale}
\end{minipage}
\hfill
\begin{minipage}{0.45\linewidth}
    \includegraphics[width=\linewidth]{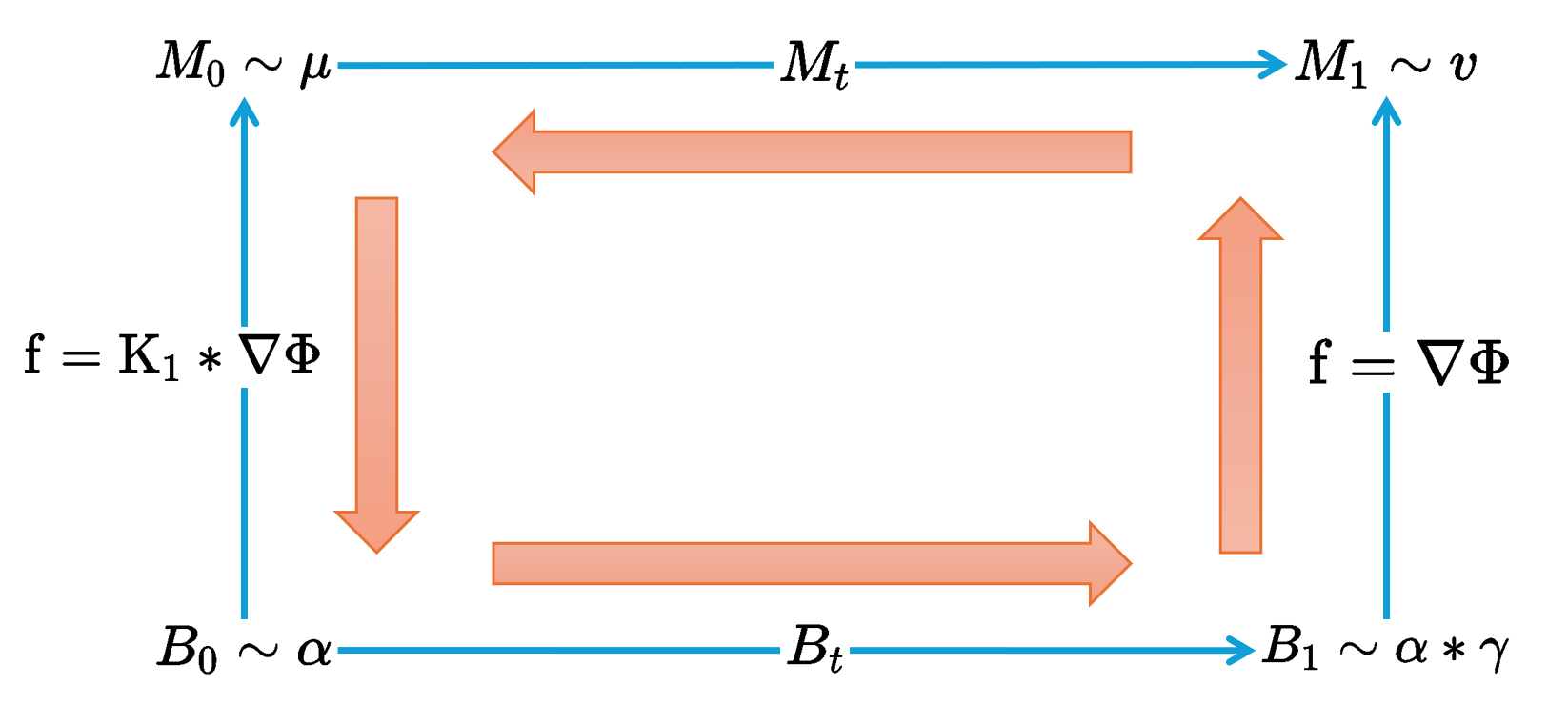}
    \caption{Bass Local Volatility}
    \label{fig:Bass Local vol}
\end{minipage}
\end{figure}

This turns out to be trivial to construct.  At the terminal, i.e. where $f$ maps from $B_1$ to $\nu$, $f$ is simply the Frechet Hoeffding solution $f=F_{\nu}^{-1}\circ F_{B_1}=F_{\nu}^{-1}\circ \mathcal{N}(\sigma)$ where $F$ and $F^{-1}$ denote the CDF and the inverse CDF of the respective distribution.  Given that the lifted space is a Martingale, $f$ must obey the heat equation.  In this vein, defining $f$ on all of $t\in[0,1]$ boils down to solving the heat equation with terminal condition defined by the Frechet Hoeffding solution.  From classical PDE theory, this is simply the convolution of the terminal condition with the heat kernel operator:
\begin{align}
    &\frac{\partial f}{\partial t}+\frac{1}{2}\frac{\partial^2f}{\partial\sigma^2}f=0,\\
    &f(\sigma,1)=F_{\nu}^{-1}\circ F_{B_1},\\
    &f(\sigma, t)=\mathcal{K}_{1-t}*f(\sigma, 1)=\mathcal{K}_{1-t}*\big(F_{\nu}^{-1}\circ F_{B_1}\big).
\end{align}
In general, when the initial marginal $\mu$ is not trivial, the base process cannot be assumed to be reversible.  Therefore, finding the optimizer for the Martingale Benamou Brenier problem is equivalent to finding the initial marginal for the base process.  Specifically, in the Bass local volatility case, the LV calibration reduces to devising a fixed point algorithm as in \citep{henry2021bass} to find the initial base distribution of $\alpha$. Figure \ref{fig:Bass Martingale} illustrates the Bass martingale, where similar idea can be seen in \citep{acciaio2023calibration}; and Figure \ref{fig:Bass Local vol} shows the main idea of Bass-LV.

The function $f_t(x): [T_i,T_{i+1}] \times \mathbb{R} \rightarrow \mathbb{R}$ defines the underlying spot price process $S_t$ between two given maturities.

To calibrate the specific $W_t$ for $t \in [T_i,T_{i+1}]$, one needs to apply a fixed-point algorithm for the cumulative distribution function (CDF) of $W_t$: $F_{W_{T_i}} = \mathcal{A} F_{W_{T_i}}$, where $\mathcal{A}$ is a nonlinear operator given by 
\[
\mathcal{A} F := F_{\mu_i} \circ\left(K_{T_{i+1}-T_i} \star\left(F^{-1}_{\mu_{i+1}} \circ\left(K_{T_{i+1}-T_i} \star F\right)\right)\right).
\]
Here, $K$ is the heat kernel $K_t(x) := \frac{e^{-\frac{x^2}{2t}}}{\sqrt{2\pi t}}$, $\circ$ denotes the composition operator, and $\star$ the convolution.

In numerical practice, we start the calibration with an initial guess of Gaussian distribution for that specific random process $W_t$. The spot price process can then be expressed by $f(t, \cdot) = K_{T_{i+1}-t} \star \left(F_{\mu_{i+1}}^{-1} \circ \left(K_{T_{i+1}-T_i} \star F_{W_{T_i}}\right)\right)$, where $F_{W_{T_i}}$ is the numerical solution of the fixed-point problem. Monte Carlo methods can then be used to price any financial option of interest by simulating the spot price process $f$.

\paragraph{Time interpolation in industry practice.}
In industry practice, generating a robust and arbitrage-free volatility surface typically begins with
fitting market quotes to a prescribed parametric representation, such as SVI \citep{gatheral2014arbitrage},
or to a discrete volatility surface as in \citep{Buehler2026sanos} and \citep{cvi2026}.
However, even after the fitted maturities are individually free of calendar and butterfly arbitrage,
the local volatility engine must still interpolate and extrapolate in the time direction at the time of pricing.
In most industry libraries, such interpolation is performed via numerical recipes such as cubic spline on implied volatilities,
which offers no structural guarantee of arbitrage freeness for the interpolated slices.

A common alternative is to interpolate the total implied variance along log-forward moneyness.
Let $\omega(t,k)$ denote the total variance at time $t$ and log-forward moneyness $k := \ln(K/F)$,
where $K$ is the absolute strike and $F$ is the forward price.
For two fitted maturities $T_1 < T_2$ and any $t \in (T_1, T_2)$, the standard linear interpolation reads
\begin{equation}\label{eq:total_var_interp}
    \omega(t,k) = \omega(T_1,k) + (t - T_1) \cdot \frac{\omega(T_2,k) - \omega(T_1,k)}{T_2 - T_1}.
\end{equation}
Under the assumption that both boundary slices are calendar-arbitrage-free,
\eqref{eq:total_var_interp} preserves the monotonicity of total variance in time and hence
the interpolated slice is guaranteed to be calendar-arbitrage-free.
However, butterfly arbitrage freeness is not guaranteed:
convexity of the call price surface in the strike direction may be violated by
the linear blending of total variances from two slices with different smile shapes.

\paragraph{Bass--LV as an arbitrage-free time interpolation scheme.}
The Bass--LV construction provides a principled alternative that simultaneously eliminates both forms of static arbitrage.
Because the construction produces a genuine martingale $(S_t)_{t\in[T_i,T_{i+1}]}$ between any two consecutive fitted maturities,
every intermediate marginal law is induced by the martingale dynamics and therefore inherits
both calendar and butterfly arbitrage freeness.
This is formalized in the following lemma.

\begin{lemma}[Arbitrage-free intermediate slices under Bass--LV interpolation]
\label{lem:bass_time_interp}
Let $(S_t)_{t\in[T_1,T_2]}$ be the Bass--LV martingale calibrated on $[T_1,T_2]$.
For any $t\in(T_1,T_2)$, define call prices
\[
C(t,K):=\mathbb{E}\big[(S_t-K)^+\big],\qquad K>0.
\]
Then, for each fixed $t$, the map $K\mapsto C(t,K)$ is decreasing and convex (hence
butterfly-arbitrage-free). Moreover, for each fixed $K$, the map $t\mapsto C(t,K)$ is nondecreasing
(hence calendar-arbitrage-free). Therefore, the implied-volatility slice obtained by Black--Scholes
inversion of $C(t,\cdot)$ is free of static arbitrage.
\end{lemma}

\noindent
The proof is deferred to Appendix~\ref{app:time_interp}.

\paragraph{End-to-end calibration-to-pricing workflow.}
The practical recipe proceeds as follows.
When a volatility surface is fitted to market quotes at maturities $\{T_i\}_{i=1}^{n}$,
we perform the Bass--LV fixed-point algorithm and store the calibrated base distributions
$\{F_{W_{T_i}}\}_{i=1}^{n}$ alongside the fitted surface.
These fixed-point solutions are reusable: once computed, any off-grid maturity
$t \in (T_i, T_{i+1})$ can be handled on demand by evaluating the Bass--LV map
\[
f(t,\cdot) = K_{T_{i+1}-t} \star \left( F_{\mu_{i+1}}^{-1} \circ \left( K_{T_{i+1}-T_i} \star F_{W_{T_i}} \right) \right),
\]
which yields the induced marginal law $\mu_t = \mathrm{Law}(S_t)$ and hence the corresponding
vanilla prices and implied volatilities without refitting the surface.
For European vanilla instruments whose maturity does not coincide with a fitted date,
this provides an immediate arbitrage-free implied volatility lookup.
For path-dependent or American style instruments priced on a time grid
(whether in a PDE or Monte Carlo engine), each grid point $t$ requires a
local volatility evaluation via Dupire's formula, which in turn requires an implied volatility
slice $\sigma(\cdot, t)$; the Bass--LV on-demand convolution supplies this slice with a structural
arbitrage-free guarantee that standard interpolation methods lack.

\section{Construction of Arbitrage-Free Risk Neutral Densities}
\label{sec:rnd}
In this section, we give details of the proposed method for obtaining arbitrage-free risk neutral density from options data. The densities are then used to generate the marginal distributions required for the construction of the Bass-LV model. We focus on formulations based on vanilla call options. Formulations based on put options can be obtained similarly.

Consider the European call option pricing formula from the Black-Scholes model:
$$
C_t\left(K, \tau\right) = S_t \Phi\left(d_1(K)\right) - K e^{-r \tau} \Phi\left(d_2(K)\right),
$$
where $S_t$ is the underlying asset price at time $t$, $K$ is the strike price, $r$ is the risk-free interest rate, $\tau = T - t$ is the time-to-maturity, and $\sigma$ is the volatility of the asset. The terms $d_1$ and $d_2$ are given by:
$$
d_1(K) = \frac{\ln \left(S_t / K\right) + \left(r + 0.5 \sigma^2\right) \tau}{\sigma \sqrt{\tau}}, \quad d_2(K) = d_1(K) - \sigma \sqrt{\tau}.
$$
By the Breeden-Litzenberger formula, we can compute the risk neutral density of the underlying asset as follows:
$$
\left.q(x) \stackrel{\text{def}}{=} e^{r \tau} \frac{\partial^2 C_t(K, \tau)}{\partial K^2}\right|_{K=x},
$$
where $q(x)$ is the risk neutral probability density of $S_T$, $C_t(K,\tau)$ is the time-$t$ price of a European call option with strike $K$ and maturity $\tau=T-t$.

Common practice to obtain the risk neutral density is to first apply some smoothing techniques to remove butterfly arbitrage in option prices, and then use the central difference method to approximate the second order derivative. However, as we will show later in the numerical experiment section, such methods will be highly sensitive to the number of market observations and will also have significant errors in reproducing the call option prices if the integral of the estimated density significantly deviates from 1.

Such numerical difficulties are not surprising considering that even small differences in prices can lead to significant variations in implied volatilities (IVs), particularly for near-maturity options. For the sake of numerical stability, instead of directly using the relationship between option prices and the risk neutral density, we express the risk neutral density as a function of the implied volatility and its derivatives, and compute the risk neutral density from IVs.

The risk neutral density as a function of implied volatility can be seen in \citep{benko2007extracting}. Here $\varphi$ is the pdf of a standard normal distribution, and $\sigma_t(K,\tau)$ is the time-$t$ implied volatility of a European vanilla call option with strike $K$ and time to maturity $\tau$:
\begin{align}
q(x) =& e^{r\tau}S_t\sqrt{\tau}\varphi\left(d_1(x)\right)\left[\frac{1}{x^2 \sigma_t(x, \tau) \tau} + \left.\frac{2 d_1(x)}{x\sigma_t(x, \tau)\sqrt{\tau}} \frac{\partial \sigma_t(K, \tau)}{\partial K}\right|_{K=x}\right.\nonumber\\
&+\frac{d_1(x) d_2(x)}{\sigma_t(x, \tau)}\left(\left.\frac{\partial \sigma_t(K, \tau)}{\partial K}\right|_{K=x}\right)^2\nonumber\\
&+\left.\left.\frac{\partial^2 \sigma_t(K, \tau)}{\partial K^2}\right|_{K=x} \right]\label{eq:spd_simga}.
\end{align}

When this formulation is used practically, we also need to impose non-arbitrage conditions. Following the settings from \citep{brunner2003arbitrage}, for a frictionless and arbitrage-free market, and for a given maturity $T$, we require the following:
\begin{enumerate}[label=\textbf{\arabic*}]
    \item Non-negativity: the risk neutral density is non-negative with $q(x) \geq 0, x \geq 0$.\label{ref:nonarbitrage condition 1}
    \item Total probability condition: the risk neutral density integrates to one, $\int_0^{\infty} q(x) \, \mathrm{d} x = 1$.\label{ref:nonarbitrage condition 2}
    \item Martingale property: the risk neutral density correctly reprices all calls, $\int_0^{\infty} \max\{x - K, 0\} q(x) \, \mathrm{d} x = \mathrm{e}^{r\tau} C_t(K, \tau), K \geq 0$.\label{ref:nonarbitrage condition 3}
\end{enumerate}

\subsection{Adaptive Local Quadratic Estimation\label{LQR}}

Following \citep{benko2007extracting}, we take observed market IVs with noise and estimate the corresponding true IVs and its derivatives. It's assumed that an observed IV consists of a true IV and some noise: $\tilde{\sigma}_i = \sigma\left(K_i\right) + \varepsilon_i$, where $\tilde{\sigma}_i$ is the observed IV, $\sigma\left(K_i\right)$ is the unknown true IV to be estimated, $K_i$ is the $i$th strike, and $\varepsilon_i$ is some unknown noise. The local quadratic estimator $\hat{\sigma}(K)$ and it's first and second order derivatives can be obtained by solving the following optimization problem,
$$
\min _{\alpha_0, \alpha_1, \alpha_2} \sum_{i=1}^{n_\tau}\left\{\tilde{\sigma}_i - \alpha_0 - \alpha_1\left(K_i - K\right) - \alpha_2\left(K_i - K\right)^2\right\}^2 \mathcal{K}_h\left(K - K_i\right),
$$
with
$$
\hat{\sigma}\left(K\right)=\alpha_0, \quad \hat{\sigma}^{\prime}\left(K\right)=\alpha_1, \quad \hat{\sigma}^{\prime \prime}\left(K\right)=2 \alpha_2.
$$
Here, $n_\tau$ is the number of available options with time-to-maturity $\tau$, $\mathcal{K}_h\left(K - K_i\right) \stackrel{\text{def}}{=} \frac{1}{h} \mathcal{K}\left(\frac{K - K_i}{h}\right)$ is a kernel function with bandwidth $h$. For example, the Epanechnikov kernel, $\mathcal{K}(u)=\frac{3}{4}\left(1-u^2\right) I(|u| \leq 1)$, is used in \citep{benko2007extracting}. Then the risk neutral density can be rewritten in terms of $\alpha_0$, $\alpha_1$ and $\alpha_2$ as below:
$$
\hat{q}(K) = F \sqrt{\tau} \varphi\left(d_1\right)\left\{\frac{1}{K^2 \alpha_0 \tau} + \frac{2 d_1}{K \alpha_0 \sqrt{\tau}} \alpha_1 + \frac{d_1 d_2}{\alpha_0}\left(\alpha_1\right)^2 + 2 \alpha_2\right\},
$$
where $\tau$ is again the time to maturity, $F=S_te^{r\tau}$ is the forward price, and $\varphi$ is the standard normal probability density function. To ensure non-negativity, we impose the following constraint on the above optimization problem:
$$
F \sqrt{\tau} \varphi\left(d_1\right)\left\{\frac{1}{K^2 \alpha_0 \tau} + \frac{2 d_1}{K \alpha_0 \sqrt{\tau}} \alpha_1 + \frac{d_1 d_2}{\alpha_0}\left(\alpha_1\right)^2 + 2 \alpha_2\right\} \geq 0.
$$

A constant bandwidth and a fixed kernel, the Epanechnikov kernel, were used in \citep{benko2007extracting}. The latter could be extended, as in \citep{fengler2015simple}, which uses a two-univariate spline kernel that can accommodate B-splines of any order. As for the former, a constant bandwidth, the choice for $h$ could influence the accuracy of the estimation significantly. The original local quadratic regression model manually chooses a constant $h$ for different market data and models. This is highly heuristic and can be unstable if we have little prior information about the market data. 

Since $K_h$ is only nonnegative within the localization window $[K - h, K + h]$, points outside of this interval have no influence on the estimate $\hat{\sigma}(K)$. When $h$ is too small, in regions with insufficient observations, not enough observations are in $[K - h, K + h]$. This leads to an estimation that is noisy and unstable. On the other hand, when $h$ is too large, the quality of estimation deteriorates due to over-smoothing. In our implementations, instead of a constant $h$, we choose $h$ adaptively so that the above window covers sufficient neighboring observations. Namely, we choose the number of points to be included in the localization window. Numerical results show that this approach leads to a more stable estimation, particularly when there are insufficient observations for extreme strikes. Adding more observations in such regions helps to achieve an estimation quality that is comparable to that in regions with abundant observations.

\subsection{Lognormal Mixture Tail Approximation}
\label{section:mixture lognormal}
In Section \ref{LQR}, we impose the non-negativity requirement \ref{ref:nonarbitrage condition 1}. Following \citep{brunner2003arbitrage}, in this section, we impose the remaining two requirements \ref{ref:nonarbitrage condition 2} \& \ref{ref:nonarbitrage condition 3} by modeling tails of the risk neutral density. We model the risk neutral density by a piecewise function:
$$
q\left(x ; \theta_L, \theta_U\right) = \begin{cases}q^L\left(x ; \theta_L\right), & x < K_L, \\ q^{\mathcal{M}}(x), & K_L \leq x \leq K_U, \\ q^U\left(x ; \theta_U\right), & x > K_U.\end{cases}
$$
Here $L, U$ stand for the lower and upper bounds of observed strikes, $\theta_L, \theta_U$ are the parameters corresponding to the left-tail and right-tail of the risk neutral density, respectively, and $q^{\mathcal{M}}(x)$ is the part of the density that is obtained in Section \ref{LQR}. It can be shown that the requirements \ref{ref:nonarbitrage condition 2} \& \ref{ref:nonarbitrage condition 3} are equivalent to the following relations:
\begin{align}
    \label{eq:log system-1}
    &\int_0^{K_L} q^L\left(x ; \theta_L\right) \mathrm{d} x + \int_{K_U}^{\infty} q^U\left(x ; \theta_U\right) \mathrm{d} x = 1 - \int_{K_L}^{K_U} q^{\mathcal{M}}(x) \mathrm{d} x,\\
    &-\left.\mathrm{e}^{r\tau} \frac{\partial C_t^{\mathcal{M}}(K, \tau)}{\partial K}\right|_{K=K_U} = \int_{K_U}^{\infty} q\left(x ; \theta_L, \theta_U\right) \mathrm{d} x,\\
    &1 + \left.\mathrm{e}^{r\tau} \frac{\partial C_t^{\mathcal{M}}(K, \tau)}{\partial K}\right|_{K=K_L} = 1 - \int_{K_L}^{\infty} q\left(x ; \theta_L, \theta_U\right) \mathrm{d} x,\\
    &F_t(\tau) = \int_0^{K_L} x q^L\left(x ; \theta_L\right) \mathrm{d} x + \int_{K_L}^{K_U} x q^{\mathcal{M}}(x) \mathrm{d} x + \int_{K_U}^{\infty} x q^U\left(x ; \theta_U\right) \mathrm{d} x\label{eq:log system-4},
\end{align}
where $F_t(\tau)$ is the forward price.

We model each tail of the risk neutral density by a mixture of two log-normal distributions, namely:
$$
q^i\left(x ; \theta_i\right) = \lambda_i \ell\left(x ; \eta_{i, 1}, v_{i, 1}^2\right) + (1 - \lambda_i) \ell\left(x ; \eta_{i, 2}, v_{i, 2}^2\right), \quad \lambda_i \in [0, 1],i \in \{L,U\},
$$
where the lognormal density function is defined as:
$$
\ell\left(x ; \eta_{i, j}, v_{i, j}^2\right) = \frac{1}{x v_{i, j} \sqrt{2 \pi}} \exp \left(-\frac{1}{2}\left(\frac{\ln(x) - \ln(\eta_{i, j}) + \frac{v_{i, j}^2}{2}}{v_{i, j}}\right)^2\right), \quad j = 1, 2; i \in \{L,U\},
$$
and $\theta_i = (\lambda_i, \eta_{i, 1}, v_{i, 1}^2, \eta_{i, 2}, v_{i, 2}^2)^\prime$ for $i \in \{L, U\}$. The reason to model each tail by a mixture of two lognormal distributions is that a single lognormal distribution is not flexible enough for us to solve the system \ref{eq:log system-1} to \ref{eq:log system-4} and meet the non-arbitrage requirements. Note that $\eta_{i,j}$ is the expected value of the corresponding lognormal distribution. If we define $\mu_{i,j}=\ln(\eta_{i,j})-\frac{1}{2}v_{i,j}^2$, then $\mu_{i,j}$ is the expected value of the corresponding normal distribution, which is more familiar.

To understand the implications of the lognormal-mixture tail completion for option prices and implied volatilities on the wings, we establish the following two results. The first shows that each mixture tail yields closed-form wing option prices; the second characterizes the asymptotic implied volatility level governed by the tail parameters.

\begin{proposition}[Closed-Form Wing Option Prices under Lognormal-Mixture Tails]
\label{proposition:wing_option_price}
For a lognormal distribution with mean parameter $\eta>0$, log-variance $v^2>0$, and probability density function $l(x; \eta, v^2)$, define
\[
d_2(K;\eta,v) := \frac{\ln(\eta/K) - \frac{1}{2}v^2}{v}, \qquad d_1(K;\eta,v) := d_2(K;\eta,v) + v.
\]
Then for any $K>0$,
\begin{align}
\int_K^\infty (x-K)\,\ell(x;\eta,v^2)\,dx &= \eta\,\Phi\!\big(d_1(K;\eta,v)\big) - K\,\Phi\!\big(d_2(K;\eta,v)\big), \label{eq:lognormal_call}\\[4pt]
\int_0^{K} (K-x)\,\ell(x;\eta,v^2)\,dx &= K\,\Phi\!\big({-d_2(K;\eta,v)}\big) - \eta\,\Phi\!\big({-d_1(K;\eta,v)}\big), \label{eq:lognormal_put}
\end{align}
where $\Phi(\cdot)$ denotes the standard normal cumulative distribution function. Given the upper tail specification $q^U(x;\theta_U) = \lambda_U \ell(x;\eta_{U,1},v_{U,1}^2) + (1-\lambda_U)\ell(x;\eta_{U,2},v_{U,2}^2)$, the price of a call on the right wing with strike price $K > K_U$ admits a closed-form expression as a convex combination of two Black-Scholes call prices:
\begin{equation}
\int_K^\infty (x-K)\,q^U(x;\theta_U)\,dx = \lambda_U\,C^{(1)}(K) + (1-\lambda_U)\,C^{(2)}(K),
\label{eq:mixture_call}
\end{equation}
where $C^{(j)}(K) := \eta_{U,j}\,\Phi\!\big(d_1(K;\eta_{U,j},v_{U,j})\big) - K\,\Phi\!\big(d_2(K;\eta_{U,j},v_{U,j})\big)$ for $j=1,2$. Similarly, given the lower tail specification $q^L(\cdot;\theta_L)$, the price of a put on the left wing with strike price $K < K_L$ admits a closed-form expression as a convex combination of two Black-Scholes put prices.
\end{proposition}

Proposition~\ref{proposition:wing_option_price} shows that deep out-of-the-money call and put option prices with strikes beyond the observed market range are fully determined by the tail completion. When the tails are specified as lognormal mixtures, these option prices admit closed-form expressions as linear combination of Black--Scholes prices. Wing implied volatilities are then obtained by inverting the Black--Scholes formula on these prices. The next result characterizes the asymptotic behavior of the implied volatility far into the wings.

\begin{proposition}[Asymptotic Wing Implied Volatility]
\label{proposition:wing_iv_asymptotics}
Under the lognormal-mixture tail model, define
\[
v_U^\ast := \max\{v_{U,1},\, v_{U,2}\}, \qquad v_L^\ast := \max\{v_{L,1},\, v_{L,2}\}.
\]
Then the implied volatility $\sigma_{\mathrm{imp}}(K,\tau)$ extracted from the wing option prices satisfies
\begin{align}
\sigma_{\mathrm{imp}}(K,\tau) &\;\longrightarrow\; \frac{v_U^\ast}{\sqrt{\tau}} \qquad \text{as } K \to \infty, \label{eq:right_wing_iv}\\[4pt]
\sigma_{\mathrm{imp}}(K,\tau) &\;\longrightarrow\; \frac{v_L^\ast}{\sqrt{\tau}} \qquad \text{as } K \to 0. \label{eq:left_wing_iv}
\end{align}
That is, the implied volatility on each wing flattens asymptotically to a level determined by the tail component with the larger log-variance.
\end{proposition}

The proofs of Propositions~\ref{proposition:wing_option_price} and~\ref{proposition:wing_iv_asymptotics} are given in Appendix~\ref{appendix:proof_wing}.

We further impose the following boundary conditions for smoothness of the risk neutral density:

\begin{align}
    \left.\frac{\partial q^L\left(x ; \theta_L\right)}{\partial x}\right|_{x=K_L} = \left.\frac{\partial q^{\mathcal{M}}(x)}{\partial x}\right|_{x=K_L} \text{ and } \left.\frac{\partial q^U\left(x ; \theta_U\right)}{\partial x}\right|_{x=K_U} = \left.\frac{\partial q^{\mathcal{M}}(x)}{\partial x}\right|_{x=K_U},\label{eq:boundary condition brunner}
\end{align}

The ten-parameter nonlinear system can be reduced to a nonlinear equation with one parameter, which can be solved numerically by standard one-dimensional root-finding methods. Details can be seen in \citep{brunner2003arbitrage}. The relationships between the parameters are:
\begin{align}
    &N\left(z_i\right) = N\left(d_2\left(K_i\right)\right) - \left.K_i n\left(d_2\left(K_i\right)\right) \sqrt{\tau} \frac{\partial \sigma_t(K, \tau)}{\partial K}\right|_{K=K_i},i \in \{L, U\},\\
    &\lambda_i = \frac{q^{\mathcal{M}}\left(K_i\right) - \frac{1}{K_i v_{i, 2} \sqrt{2 \pi}} \mathrm{e}^{-\frac{1}{2} z_i^2}}{\frac{1}{K_i v_{i, 1} \sqrt{2 \pi}} \mathrm{e}^{-\frac{1}{2} z_i^2} - \frac{1}{K_i v_{i, 2} \sqrt{2 \pi}} \mathrm{e}^{-\frac{1}{2} z_i^2}},\\
    &\eta_{i, 1} = K_i \mathrm{e}^{z_i v_{i, 1} + \frac{v_{i, 1}^2}{2}}, \quad \eta_{i, 2} = K_i \mathrm{e}^{z_i v_{i, 2} + \frac{v_{i, 2}^2}{2}},\\
    &v_{i, 1} = \frac{q^{\mathcal{M}}\left(K_i\right) - \frac{1}{K_i v_{i, 2} \sqrt{2 \pi}} \mathrm{e}^{-\frac{1}{2} z_i^2}}{\left(q^{\mathcal{M}}\left(K_i\right) + \left.K_i \frac{\partial q^{\mathcal{M}}}{\partial x}\right|_{x=K_i}\right) \frac{1}{z_i} - \frac{q^{\mathcal{M}}\left(K_i\right)}{v_{i, 2}}},\\
    & \lambda_L \eta_{L, 1} N\left(-z_L-v_{L, 1}\right)+\left(1-\lambda_L\right) \eta_{L, 2} N\left(-z_L-v_{L, 2}\right) \nonumber\\
    & =F_t(\tau) N\left(-d_1\left(K_L\right)\right)+\left.K_L^2 n\left(d_2\left(K_L\right)\right) \sqrt{\tau} \frac{\partial \sigma_t(K, \tau)}{\partial K}\right|_{K=K_L}, \\
    & \lambda_U \eta_{U, 1} N\left(z_U+v_{U, 1}\right)+\left(1-\lambda_U\right) \eta_{U, 2} N\left(z_U+v_{U, 2}\right)\nonumber \\
    & =F_t(\tau) N\left(d_1\left(K_U\right)\right)-\left.K_U^2 n\left(d_2\left(K_U\right)\right) \sqrt{\tau} \frac{\partial \sigma_t(K, \tau)}{\partial K}\right|_{K=K_U}.
\end{align}
Here $n(\cdot)$ is the standard normal pdf and $N(\cdot)$ is the standard normal cdf.

Finally, to prevent calendar arbitrage, we require that the risk neutral densities for different maturities satisfy
\begin{equation}
\int_{0}^{\infty} \max\{x - K, 0\} \left( e^{r(T_{i+1} - T_i)} q_{S_{T_{i+1}}}(x) - q_{S_{T_i}}\left( x e^{-r(T_{i+1} - T_i)} \right) \right) dx \geq 0,
\label{eq:calendar_arbitrage_condition}
\end{equation}
for all maturities \(\ T_i \leq T_{i+1} \). 

\citet{brunner2003arbitrage} formulate the tail completion for a single maturity and leave the liquid-region estimation as an unspecified, market-dependent input.
Our objective is to unify the middle-region estimation and the tail completion into a coherent, fully specified marginal construction that serves as a reliable downstream component of the Bass--LV fixed-point iterations. In Bass--LV, the CDF and inverse-CDF are repeatedly evaluated at every iteration, so global computability and numerical stability become first-order requirements rather than secondary smoothness preferences.

To this end, in addition to imposing Conditions
\ref{ref:nonarbitrage condition 1}--\ref{ref:nonarbitrage condition 3}
at each maturity, we enforce the boundary and monotonicity constraints in Eq.~\eqref{eq:boundary condition brunner} at the middle--tail stitching points.  The matching order of these constraints is not chosen for cosmetic smoothness alone: it determines the effective
smoothness index~$m$ that enters the trapezoidal-rule convergence rate in Section~\ref{sec: optimality theory}, and therefore must be set in conjunction with the optimal parameter selection for the downstream convolution scheme.  This co-design ensures that the constructed
density, CDF, and inverse-CDF remain stable under repeated evaluation inside the fixed-point solver. Moreover, repeated downstream querying
can expose numerical fragility in the tail-completion system itself:
in Section~\ref{sec:tsla}, a near-zero mode implied by the middle region
makes the nonlinear tail system ill-conditioned. We report reproducible
safeguards, consisting of a slight domain adjustment and mild
tail-parameter constraints, that restore tractability.

\section{Optimality and Convergence of Numerical Convolution}
\label{sec: optimality theory}

In this section, we demonstrate optimal numerical parameter selection and convergence of the trapezoidal rule for numerical convolution in Bass-LV implementation. In particular, we assume limited smoothness in the marginal distribution functions and their inverses, as commonly encountered in numerical implementations involving spline interpolation and extrapolation.

Before analyzing numerical convergence, let us recall the following operator in the Bass-LV model:
$$\mathcal{A} F := F_{\mu_i} \circ \left(K_{T_{i+1}-T_i} \star \left(F_{\mu_{i+1}}^{-1} \circ \left(K_{T_{i+1}-T_i} \star F\right)\right)\right).$$ 

Here, $K$ represents the heat kernel, $\star$ denotes convolution, and $\circ$ is function composition. The analysis of this operator's properties relies on results from \citep{henry2021bass}, which we now summarize:

Lemma 2.2 in \citep{henry2021bass} establishes that $\mathcal{A}(\mathcal{E}) \subset \mathcal{E}$, where $\mathcal{E}$ is the space of cumulative distribution functions. This ensures that each iteration produces a well-defined CDF.

Furthermore, Theorem 2.4 in \citep{henry2021bass} demonstrates that $\mathcal{A}(\mathcal{E})$ is uniformly bounded and Lipschitz. By the Arzelà-Ascoli theorem, this implies that $\mathcal{A}$ is continuous in the sup-norm and $\mathcal{A}(\mathcal{E})$ is relatively compact. Since $\mathcal{E}$ is convex and closed, Schauder's fixed point theorem guarantees the existence of a solution in $\mathcal{E}$ for the fixed point problem $F=\mathcal{A} F$. This solution is the $F_{W_{T_i}}$ that we are seeking in the Bass-LV implementation.

Given two consecutive maturities $T_i, T_{i+1}$, the following multi-layer integration needs to be done in solving the fixed-point problem:
\begin{align}
& F_{\mu_i}\left(\left(K_{T_{i+1}-T_i} \star F_{\mu_{i+1}}^{-1}\left(K_{T_{i+1}-T_i} \star F\right)\right)(w)\right) \nonumber\\
= & F_{\mu_i}\left(\int_{-\infty}^{\infty} \rho(y) \cdot (F_{\mu_{i+1}}^{-1}(K_{T_{i+1}-T_i} \star F))(w-y) \, dy\right) \nonumber\\
= & F_{\mu_i}\left(\int_{-\infty}^{\infty} \rho(y) \cdot F_{\mu_{i+1}}^{-1}\left(\int_{-\infty}^{\infty} F(w-y-x) \rho(x) \, dx\right) \, dy\right).\label{eq:double integrand}
\end{align}
where $\rho(\cdot)$ is given by the heat kernel $K_{T_{i+1}-T_i}$, and $F(w)$ is the distribution obtained in a previous iteration that, upon convergence of the algorithm, yields the final distribution $F_{W_{T_i}}$.

In practice, the marginal distributions $F_{\mu_i}$, its inverse $F_{\mu_i}^{-1}$, and the intermediate iterate $F$ are never given in closed form.  Instead, they are computed on a finite grid and interpolated (and, where necessary, extrapolated) with cubic or other splines. Our theoretical analysis is established based on such limited smoothness. To be specific, the smoothness order of a function refers to the highest order of continuous derivatives it possesses. A function with smoothness order
$m$ (denoted as $C^m$) has continuous derivatives up to order $m$. 

Because splines are $C^m$ for some finite $m$, the functions that actually enter~\eqref{eq:double integrand} are automatically also $C^m$. In the analysis below, we make the following working assumption:

\begin{assumption}
\label{assumption:regularity}
$F_{\mu_i}, F_{\mu_i}^{-1}, F$ are piecewise polynomial and belong to  $C^m$ for some fixed and finite $m$.
\end{assumption}


Note that this assumption ensures smoothness at the knots between different polynomial segments while guaranteeing that the functions as a whole are in $C^m$.With this assumption, we have the following proposition:

\begin{proposition}
\label{proposition:func in weighted Sobolev Space}
Given assumption \ref{assumption:regularity}, both the internal and external integrands in Eq.\ref{eq:double integrand} are well-defined in the following weighted Sobolev space
$$
\mathscr{H}_m:=\left\{f \in L_\rho^2 \ \ \Big| \ \ \|f\|_m:=\left(\sum_{\tau=0}^m\left\|f^{(\tau)}\right\|_{L_\rho^2}^2\right)^{1 / 2}<\infty\right\},
$$
where $\rho(x)=\frac{1}{\sqrt{2 \pi \sigma^2}} e^{-\frac{x^2}{2 \sigma^2}}$, $\sigma^2=T_{i+1}-T_i$, $L_\rho^2:=L_\rho^2(\mathbb{R})$, $f^{(\tau)} \in L_\rho^2$ for $\tau=1, \ldots, m$. The convergence rate of Gauss-Hermite quadrature for the above integrals can achieve \(\mathcal{O}\left(n^{-m / 2}\right)\), where \(n\) stands for the number of quadrature points, \(m\) stands for the order of smoothness for the integrands.
 
\end{proposition}

In the following, we present results for the optimal step size and truncation level when the trapezoidal rule is used for the above numerical convolution, and give the corresponding convergence rate. We first give the trapezoidal scheme for the integral in Eq.\ref{eq:double integrand} as follows:
\begin{align}
    \sum_{m=-M}^{M} \rho(mh) \cdot F_{\mu_{i+1}}^{-1}\left(\sum_{n=-N}^{N} F(w - mh - nh) \rho(nh)h\right) \cdot h,
\end{align}
where $h$ is step size, $M$ and $N$ represent the number of terms used in the trapezoidal scheme, and $Mh$ and $Nh$ are the level of truncation when integrals on the whole real line are replaced by integrals on finite intervals. 

Following the idea of proposition 4.2 in \citep{kazashi2023suboptimality}, we can derive the following optimal setting for the step size and truncation level for trapezoidal rule based numerical convolution in Eq.\ref{eq:double integrand}:

\begin{proposition}
\label{proposition:opt trap}
Given assumption \ref{assumption:regularity}. Let $\epsilon \in (max\{1-\sigma^2,0\},1)$, where $\sigma^2=T_{i+1}-T_{i}$. We have the following optimal parameter setting for the internal integral:
\begin{align*}
    Nh &= \sqrt{\frac{2(T_{i+1}-T_{i})}{(1-\epsilon)} m \ln(2N + 1)}, \\
    h &= \frac{\sqrt{\frac{2(T_{i+1}-T_{i})}{(1-\epsilon)} m \ln(2N + 1)}}{N}.
\end{align*}
Optimal parameter setting for the external integral is similar. Under these optimal settings, the convergence rate of the trapezoidal rule for the above integrals can achieve \(\mathcal{O}\left(\frac{(\ln n)^{(m / 2+1 / 4)}}{n^m}\right)\), where $n$ stands for the number of terms used in the trapezoidal schemes, and $m$ stands for the order of smoothness for the integrands.
\end{proposition}

We conclude that, for any given convolution integral, when the same number of points are used, the convergence rate of the trapezoidal rule is \(\mathcal{O}\left(\frac{(\ln n)^{(m / 2+1 / 4)}}{n^m}\right)\) which is faster than the Gauss Hermite quadrature with a rate of \(\mathcal{O}\left(n^{-m / 2}\right)\). This provides a theoretical foundation for using the trapezoidal rule for the numerical convolutions in the implementation of the Bass-LV model.

Together with the robust marginal construction in Section 3 (including the mix lognormal tail completion adapted with additional boundary/monotonicity and calendar-consistency checks), this yields a numerically stable and computationally efficient pipeline for Bass-LV calibration in our setting.

\section{Numerical Experiments}
\label{sec:lab}
\subsection{Step-by-Step Calibration Procedure}

In this section, we outline a step-by-step procedure for calibrating the Bass-LV model from European vanilla options market data. 

\begin{enumerate}
    \item \textbf{Extract implied volatilities (IV) from market prices:} Begin with  observed market prices of European vanilla options. Compute the implied volatilities for different strikes and maturities.
    \item \textbf{Fit a local quadratic regression model:} Use the extracted IVs to fit a local quadratic regression (LQR) model. This involves solving the following optimization problem for $\alpha_0, \alpha_1,$ and $\alpha_2$:
    \[
    \min _{\alpha_0, \alpha_1, \alpha_2} \sum_{i=1}^{n_\tau}\left\{\tilde{\sigma}_i - \alpha_0 - \alpha_1\left(K_i - K\right) - \alpha_2\left(K_i - K\right)^2\right\}^2 \mathcal{K}_h\left(K - K_i\right),
    \]
    where $\mathcal{K}_h$ is a kernel function with bandwidth $h$.
    \item \textbf{Calculate the risk neutral density:} Using $\alpha_0, \alpha_1,$ and $\alpha_2$ obtained above, compute the risk neutral density $q^\mathcal{M}(x)$ on the interval $\mathcal{M}=[K_L, K_U]$ as follows:
    \[
    q^\mathcal{M}(x) = F \sqrt{\tau} \varphi\left(d_1\right)\left\{\frac{1}{x^2 \alpha_0 \tau} + \frac{2 d_1}{x \alpha_0 \sqrt{\tau}} \alpha_1 + \frac{d_1 d_2}{\alpha_0}\left(\alpha_1\right)^2 + 2 \alpha_2\right\},
    \]
    Here, $K_L$ and $K_U$ refer to the minimum and maximum observed strikes from the market data, respectively. After completing this step, one obtains the portion of the risk neutral density corresponding to the market observations.
    
    \item \textbf{Construct the tails of the risk neutral density:} Use mixture of lognormal distributions to construct tails of the risk neutral density. This can be achieved by leveraging the parameters obtained in the previous step and then solving the root-finding system in section \ref{section:mixture lognormal}.
    After completing this step, one obtains the risk neutral density on $[0, K_L]$ and $[K_U, \infty]$.
    
    \item \textbf{Complete the risk neutral density:} Combine the densities from the steps above to form a complete risk neutral density $q(x; \theta_L, \theta_U)$:
    \[
    q(x ; \theta_L, \theta_U) = \begin{cases}q^L(x ; \theta_L), & x < K_L \\ q^{\mathcal{M}}(x), & K_L \leq x \leq K_U \\ q^U(x ; \theta_U), & x > K_U\end{cases}.
    \]
    \item \textbf{Calibrate the Bass-LV model:} With the arbitrage-free risk neutral densities and the corresponding distributions obtained above as input, using trapezoidal numerical convolution, solve the fixed-point problem to perform the calibration. It involves iteratively calculating the following:
    \[
    \mathcal{A} F := F_{\mu_i} \circ \left(K_{T_{i+1}-T_i} \star \left(F_{\mu_{i+1}}^{-1} \circ \left(K_{T_{i+1}-T_i} \star F\right)\right)\right),
    \]
    until convergence is achieved. 
\end{enumerate}

This step-by-step procedure provides a structured approach to calibrate the Bass-LV model. 

\subsection{Iteration and Calibration Errors}

In the Bass construction, a fix-point problem needs to be solved iteratively. Following the practice in \citep{henry2021bass}, the stopping condition is specified by controlling the following iteration error measured in the infinite norm:
$$err_{itr}=\left\|F_{W_{T_i}}^{(p)}-F_{W_{T_i}}^{(p-1)}\right\|_{\infty}.$$
The iteration continues until the iteration error is less than a predetermined tolerance level. The iteration error tolerance naturally determines the quality of calibration. Let $err_{cab}$ be the calibration error, which is the mean absolute percentage error of calibrated IVs:
$$
    err_{cab}=\frac{1}{L}\sum_{j=1}^L\left|\frac{IV_{cab}^j-IV_{True}}{IV_{True}}\right|.
$$
Here $L$ is the number of option strike prices considered. In the following experiments, we numerically investigate how calibration error depends on iteration error tolerance.

\subsection{Experiments in the Black-Scholes-Merton Case}

In this section, we examine the Bass-LV calibration in the Black-Scholes-Merton model. In this case, the marginal distributions and their inverse are known and hence do not introduce any implementation error. The exact solution of the fixed point problem is also available in closed-form. This allows us to examine how iteration error control in numerical solution of the fixed point problem impacts calibration performance. It also enables us to compare different integration schemes and highlight the advantages of trapezoidal numerical convolution. 

Let $\mu_1, \mu_2$ and $\mu_3$ be lognormal distributions, where the standard deviations of the corresponding normal distributions are 
$\sigma \sqrt{T_1}, \sigma \sqrt{T_2}$ and $\sigma \sqrt{T_3}$, respectively. In this case, the solution to the fixed point problem is $F_{W_{T_i}}=\mathcal{N}\left(\frac{\cdot}{\sqrt{T_i}}\right)$ and $f(t, w)=S_0 \exp \left(-\frac{1}{2} \sigma^2 t+\sigma w\right)$. 

In our experiment, the current time is $T_0=0$. Options with the following maturities are considered: $T_1=1, T_2=1.2$ and $T_3=1.5$. The initial asset price is $S_0=100$, the true IV is $\sigma=1$, and the risk-free interest rate is $r=0$. The details of the experiment are given in Algorithm 1 below.

\begin{algorithm}
\caption{The Black-Scholes-Merton Case}
\begin{algorithmic}
\STATE Step 1: Given the lognormal marginal distributions: $S_{T 1} \sim \mu_1, S_{T 2} \sim \mu_2, S_{T 3} \sim \mu_3$
\STATE Step 2: Numerically solve the fixed-point problem to get $F_{W_{T_i}}$
\STATE Step 3: Simulate the spot price process and estimate European call option prices
\STATE Step 4: Compute the IVs associated with the above call prices and compare to the true IV
\end{algorithmic}
\end{algorithm}

\begin{figure}[ht]
\centering
\begin{minipage}{0.45\linewidth}
    \includegraphics[width=\linewidth]{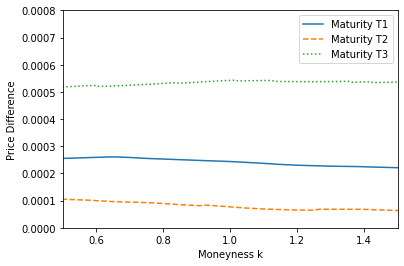}
    \caption{Pricing errors of options with maturities $T_1=1$, $T_2=1.2$, $T_3=1.5$ and moneyness $k=K/S_0$ in the calibrated Bass-LV model. Iteration error tolerance $=10^{-5}$. }
    \label{fig:priceDiff}
\end{minipage}
\hfill
\begin{minipage}{0.5\linewidth}
    \includegraphics[width=\linewidth]{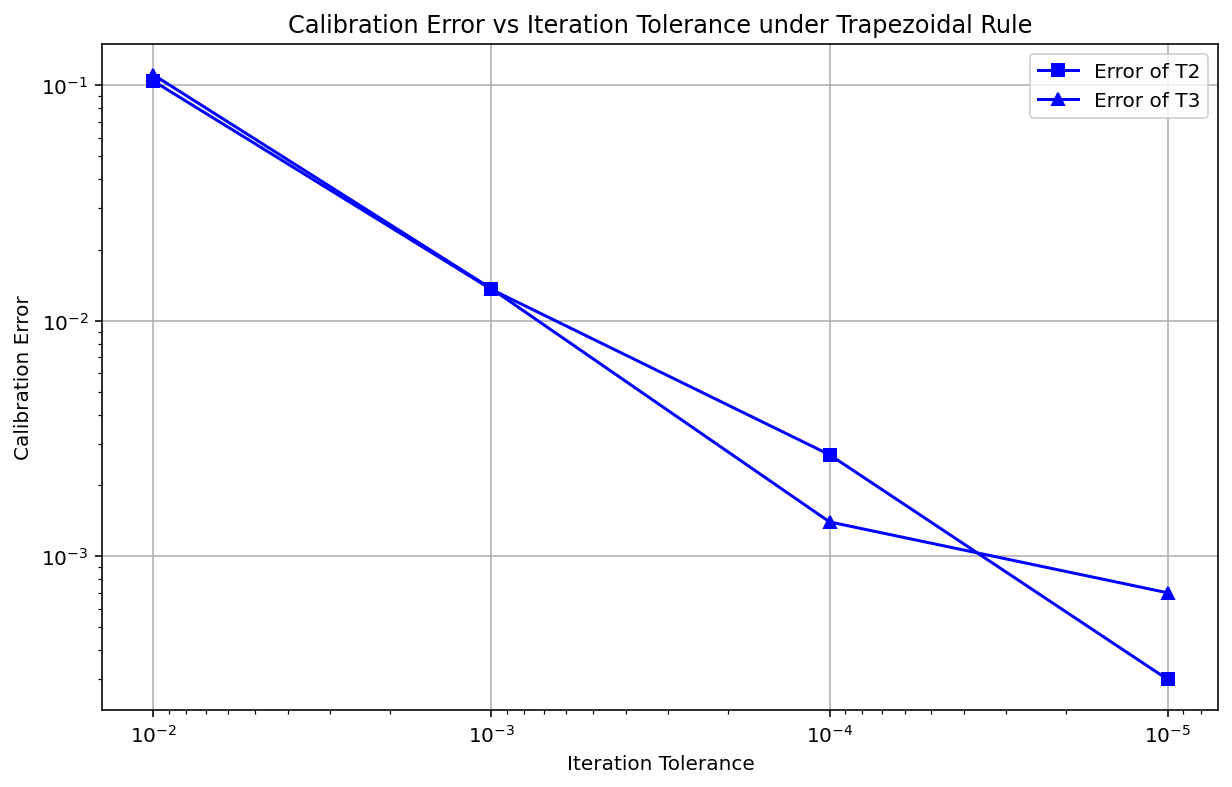}
    \caption{MAPE of calibrated implied volatility in the Black-Scholes-Merton case. Trapezoidal rule used for numerical integration.}
    \label{fig:MAE under Trapezoidal Rule of BS model}
\end{minipage}
\end{figure}

\begin{figure}[ht]

\centering
\begin{minipage}{0.7\linewidth}
    \includegraphics[width=\linewidth]{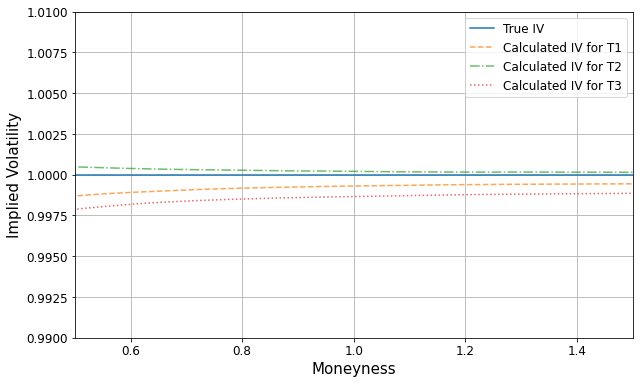}
    \caption{Calibrated IV in the Black-Scholes-Merton case for maturities $T_1=1$, $T_2=1.2$, $T_3=1.5$ and moneyness $k=K/S_0$.}
    \label{fig:Calibrated IV under Black-Scholes Model for Different Maturities}
\end{minipage}
\end{figure}

Figure \ref{fig:priceDiff} shows that the absolute pricing errors of European call options with the above three maturities and various strikes are bounded by $6\times 10^{-4}$ in the calibrated model. Here, we consider option strike prices that are in $[0.5S_0,1.5S_0]$. Options with these strikes are most actively traded. Figure \ref{fig:Calibrated IV under Black-Scholes Model for Different Maturities} shows that the implied volatilities of the above options in the calibrated Bass-LV model are very close to the true IV, which is 1. 

Iteration error tolerance in Figures \ref{fig:priceDiff} and \ref{fig:Calibrated IV under Black-Scholes Model for Different Maturities} is set to \(10^{-5}\), i.e, the fixed point algorithm will stop when \(\left\|F_{W_{T_i}}^{(p)}-F_{W_{T_i}}^{(p-1)}\right\|_{\infty}\leq 10^{-5}
\) for some iteration index $p$.

In Figure \ref{fig:MAE under Trapezoidal Rule of BS model}, we show the numerical relationship between iteration error tolerance and calibration error. European option prices are computed using Monte Carlo simulation with sample size $3 \times 10^8$. Calibrated IVs are then computed from these estimated option prices. When plotting $err_{cab}$, we re-run the Monte Carlo method several times with different random seeds and plot the median $err_{cab}$. This is to mitigate the impact of noisy random number generators and show a clearer relationship between iteration error tolerance and calibration accuracy. 
Since no fixed point problem needs to be solved for options with maturity $T_1$, we didn't plot the error for this maturity.

As can be seen in Figure \ref{fig:MAE under Trapezoidal Rule of BS model}, as the iteration error tolerance decreases, calibration accuracy improves approximately linearly. In this particular example, to get a calibration error of 1\%, an iteration error tolerance of about $10^{-3}$ is needed. In numerical experiments in \citep{henry2021bass}, an iteration error tolerance of $2 \times 10^{-3}$ was used. However, when higher accuracy levels are desired in some applications, one must use a much smaller iteration error tolerance. For example, if the desired calibration error is 0.01\%, an iteration error tolerance of $10^{-5}$ would be needed. In such cases, trapezoidal rule based numerical integration schemes clearly become more advantageous compared to the Gauss-Hermite quadrature, as to be shown next. Note that the calibration error for maturity $T_3$ flattens 
after reaching $10^{-3}$. This is due to not large enough Monte Carlo sample size. Consequently, error due to Monte Carlo estimation starts to dominate. Increasing Monte Carlo sample size or using variance reduction techniques will help further reducing the calibration error.

\begin{figure}[ht]
\centering
\begin{minipage}{0.4\linewidth}
    \includegraphics[width=\linewidth]{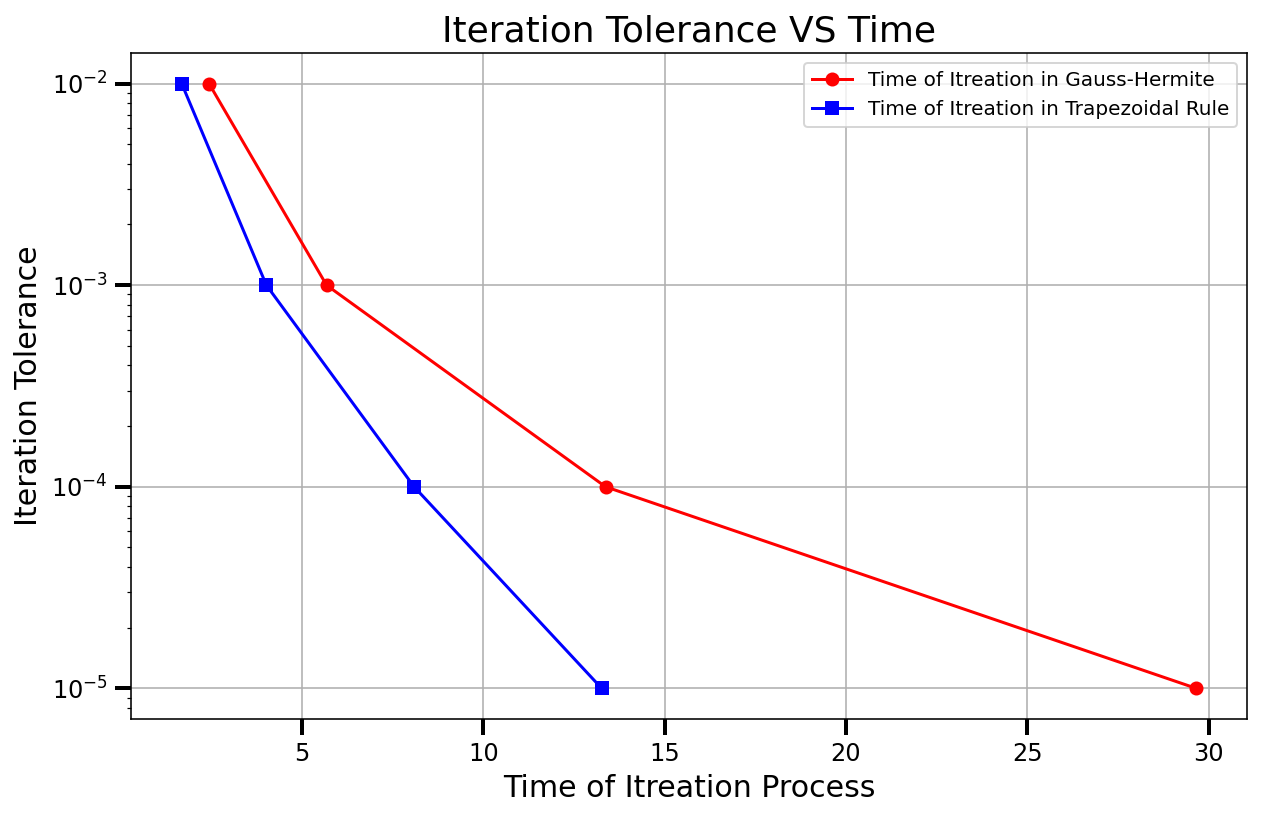}
    \caption{Time for solving fixed point problems in the Black-Scholes-Merton model: trapezoidal rule vs. Gauss-Hermite quadrature}
    \label{fig:Time compare bs}
\end{minipage}
\hfill
\begin{minipage}{0.5\linewidth}
    \includegraphics[width=\linewidth]{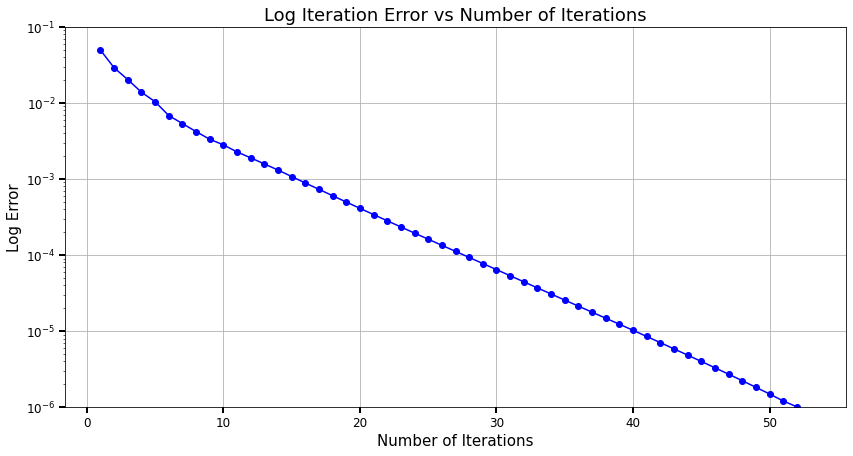}
    \caption{Iteration error tolerance vs. number of iterations needed to solve the fixed point problem in the Black-Scholes-Merton model}
    \label{fig:Iteration Tolerance vs. Iteration Numbers under the Black-Scholes Model}
\end{minipage}
\end{figure}

Figure \ref{fig:Time compare bs} compares the performance of Gauss-Hermite quadrature and the trapezoidal rule under different iteration error tolerance settings. The horizontal axis is the minimal amount of time the numerical solution of the fixed point problem takes to achieve the smallest possible calibration accuracy. It shows that, as the iteration error tolerance decreases, the computational time required when using the trapezoidal rule becomes much smaller compared to Gauss Hermite quadrature. This clearly shows the advantage of using the trapezoidal rule, in particular, when the desired accuracy level is high.

Our numerical results also support the linear convergence theory of the Bass calibration process presented in \citep{acciaio2023calibration}. Figure \ref{fig:Iteration Tolerance vs. Iteration Numbers under the Black-Scholes Model} clearly shows that,  for a given maturity, the number of iterations needed to solve the fixed point problem grows linearly in the logarithm of the iteration error tolerance. In later numerical experiments, we show that this is also roughly true in much more general settings than those specified in \citep{acciaio2023calibration}.

Tables \ref{tab:Results from Gauss Hermite Quadrature under BS case} and \ref{tab:Results from Trapezoidal Rule Scheme under BS case} provide detailed results for the plots. These experiments were conducted using a 14th Gen Intel(R) Core(TM) i9-14900HX CPU @ 2.20 GHz and Python 3.11.9. All subsequent experiments are performed within this environment. The first column contains the iteration error tolerance. The second column contains the minimal amount of time needed to solve the fixed point problem to achieve the smallest possible calibration error. For the third column, note that no fixed point problem is solved for the first maturity. The error here is totally due to Monte Carlo estimation, hence identical for different iteration error tolerance levels. The last two columns contain the calibration errors for the remaining two maturities.  

\begin{table}[H]
\centering
\caption{Time and calibration error in the Black-Scholes-Merton model: Gauss Hermite quadrature}
\label{tab:Results from Gauss Hermite Quadrature under BS case}
\scalebox{1}{
\begin{tabular}{ccccc}
\toprule
Iteration  & Iteration  & Calibration  & Calibration & Calibration  \\
tolerance & time (seconds)        & error ($T_1$) & error ($T_2$) & error ($T_3$) \\
\midrule
1.0e-2 & 2.4 & 8.0e-4 & 1.1e-1 & 1.1e-1 \\

1.0e-3 & 5.7 & 8.0e-4 & 1.3e-2 & 1.4e-2 \\

1.0e-4 & 13.4 & 8.0e-4 & 2.3e-3 & 2.1e-3 \\

1.0e-5 & 29.7 & 8.0e-4 & 3.0e-4 & 6.0e-4 \\
\bottomrule
\end{tabular}}
\end{table}

\begin{table}[H]
\centering
\caption{Time and calibration error in the Black-Scholes-Merton model: trapezoidal rule}
\label{tab:Results from Trapezoidal Rule Scheme under BS case}
\scalebox{1}{
\begin{tabular}{cccccc}
\toprule
Iteration  & Iteration  & Calibration  & Calibration & Calibration  \\
tolerance & time (seconds)        & error ($T_1$) & error ($T_2$) & error ($T_3$) \\
\midrule
1.0e-2 & 1.7  &  8.0e-4 & 1.0e-1 & 1.1e-1 \\
1.0e-3 & 4.0  &  8.0e-4 & 1.4e-2 & 1.4e-2 \\
1.0e-4 & 8.1  &  8.0e-4 & 2.7e-3 & 1.4e-3 \\
1.0e-5 & 13.3 &  8.0e-4 & 3.0e-4 & 7.0e-4 \\
\bottomrule
\end{tabular}}
\end{table}

\subsection{Comparison with the Breeden-Litzenberger Approach}
Generating the risk neutral density is one of the most important steps in Bass LV calibration. In this section, we compare our proposed method with the widely used Breeden-Litzenberger formula. The Breeden-Litzenberger approach usually starts with cleaning and smoothing the market prices to meet non-arbitrage conditions. The risk neutral density is then derived using the formula:
$$
    \left.q(x) \stackrel{\text { def }}{=} e^{r \tau} \frac{\partial^2 C_t(K, \tau)}{\partial K^2}\right|_{K=x}.
$$

To eliminate the need for the non-trivial price cleaning and smoothing step for the Breeden-Litzenberger approach, we generate arbitrage-free option prices and use them as "market" data. The experiment is done in  the Stochastic Volatility Inspired (SSVI) model from \citep{gatheral2014arbitrage}. 
Given the arbitrage-free option price surface generated in this model, we apply the Breeden-Litzenberger formula to derive the risk neutral densities for different maturities using finite difference.

Our proposed method on the other hand is rather robust, even when market data contains noise. To numerically illustrate this, we add some random noise to the arbitrage-free SSVI surface. This produces a pseudo-market IV surface with potential arbitrage. We then apply our proposed method to generate risk neutral densities and corresponding marginal distributions for different maturities. 
An outline of the experiments in the SSVI case is in Algorithm 2. 

\begin{algorithm}
\caption{The SSVI Case}
\begin{algorithmic}
\STATE Step 1: Generate "market" data from the SSVI model. Add random noise. Apply the LQR method to construct the central part of the risk neutral density.
\STATE Step 2: Use lognormal mixture to construct the tails of the risk neutral density.
\STATE Step 3: Numerically solve the fixed-point problem to get $F_{W_{T_i}}$
\STATE Step 4: Simulate the spot price process and estimate European call option prices
\STATE Step 5: Compute the IVs associated with the above call prices and compare to the true IVs.
\end{algorithmic}
\end{algorithm}

Recall that for the Heston-like SSVI model presented in \citep{gatheral2014arbitrage}, the function $\varphi$ is defined as:
$$
\varphi(\theta_t)=\frac{1}{\lambda \theta_t}\left\{1-\frac{1-\mathrm{e}^{-\lambda \theta_t}}{\lambda \theta_t}\right\},
$$
where $\lambda \geq\frac{(1+|\rho|)}{4}$ ensures no arbitrage. The total variance surface is given by:
$$
w\left(k, \theta_t\right)=\frac{\theta_t}{2}\left\{1+\rho \varphi\left(\theta_t\right) k+\sqrt{\left(\varphi\left(\theta_t\right) k+\rho\right)^2+\left(1-\rho^2\right)}\right\}.
$$
The corresponding IV surface is then obtained via
\begin{align}
\sigma(k, t)=\sqrt{\frac{w\left(k, \theta_t\right)}{t}},
\label{eq:spd_sigma}
\end{align}
where $k$ represents log moneyness, i.e.
$$
e^k=\frac{K}{S_0 e^{r \tau}},
$$
with $\tau$ being the time to maturity and $K$ the strike price.

In this model, the risk neutral density can be computed analytically. This allows us to examine the quality of the risk neutral densities estimated using both the proposed method and the Breeden-Litzenberger approach. More specifically, by computing $\sigma$, $\sigma'$, $\sigma''$ and using equation 
\ref{eq:spd_simga}, we can obtain the closed-form expression for the true risk neutral density to serve as a benchmark. 

\begin{figure}[ht]
\centering
\begin{minipage}{0.42\linewidth}
\includegraphics[width=\linewidth]{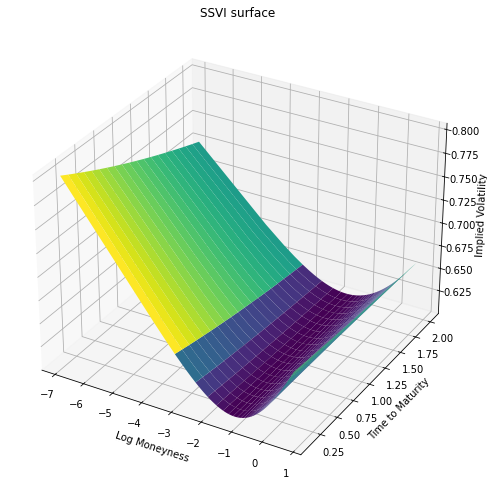}
    \caption{Heston-like SSVI implied volatility surface}
    \label{fig:heston-like-ssvi}
\end{minipage}
\hfill
\begin{minipage}{0.45\linewidth}
\includegraphics[width=\linewidth]{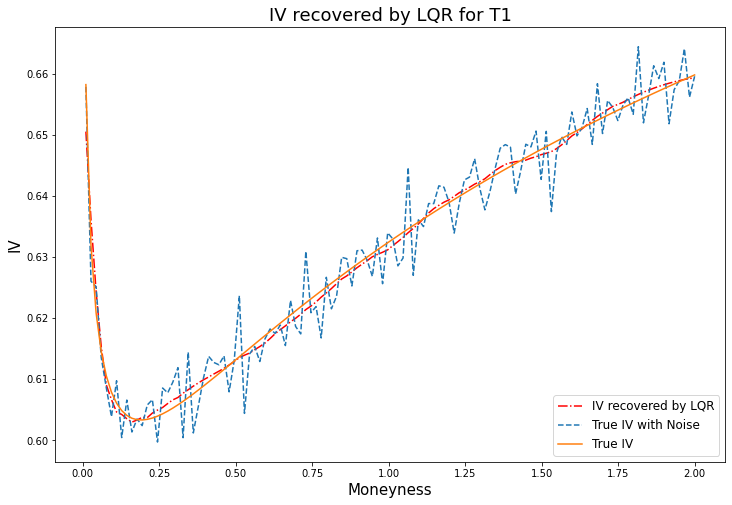}
    \caption{IVs recovered by local quadratic regression for maturity $T_1$ from data with noise}
    \label{fig:combined-opt}
\end{minipage}
\end{figure}

For our experiment, we set the parameters as follows: $\rho = 0.3$, $\lambda = \frac{(1+|\rho|)}{4} + 1$, and $\theta_t = 0.4t$. The initial spot price is set at $S_0 = 100$. The risk-free rate is $r = 0$. Figure \ref{fig:heston-like-ssvi} illustrates the resulting SSVI surface. 

\begin{figure}[ht]
\centering
\begin{minipage}{0.45\linewidth}
\includegraphics[width=\linewidth]{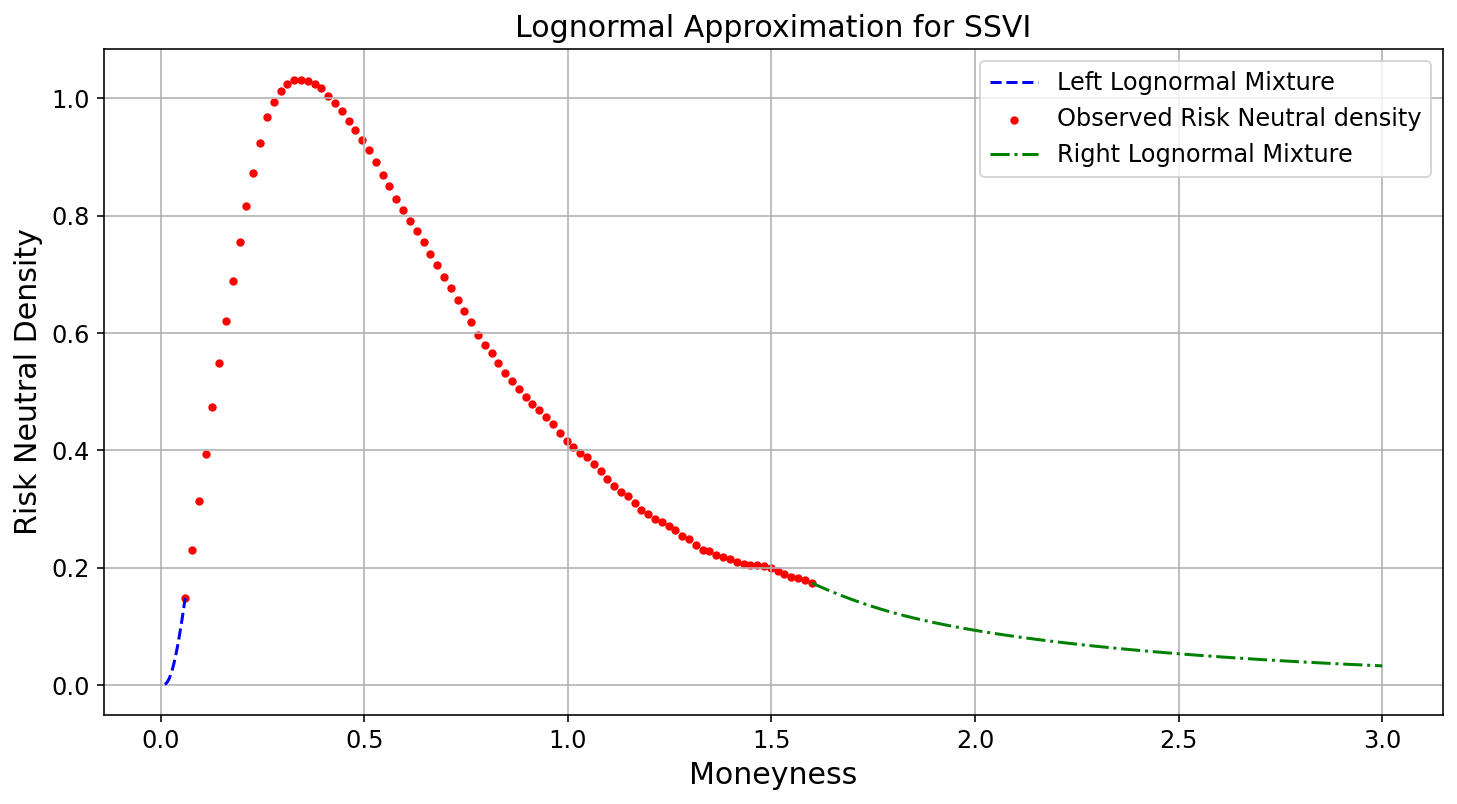}
    \caption{Estimated risk neutral density with lognormal-mixture tails from data with noise}
    \label{fig:Lognormal Mixture Approximation under Pasting}
\end{minipage}
\hfill
\begin{minipage}{0.45\linewidth}
\includegraphics[width=\linewidth]{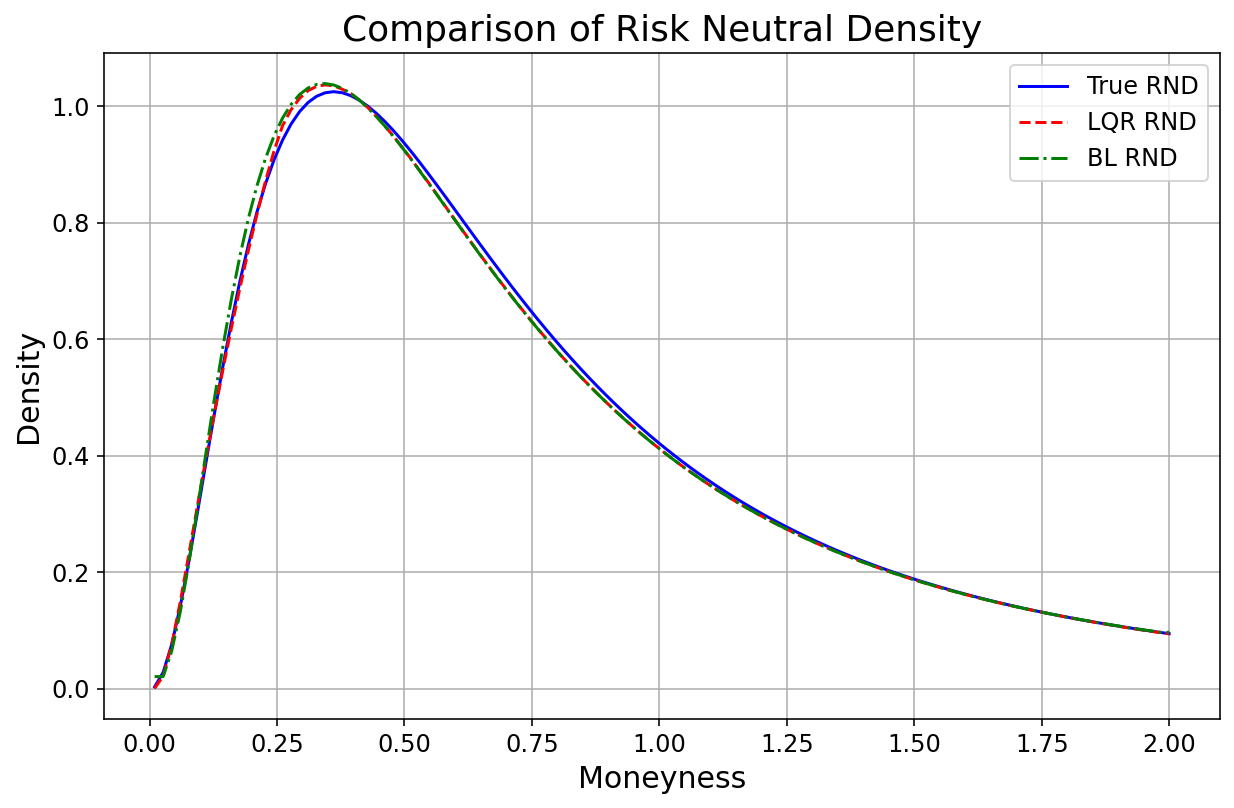}
    \caption{Estimated risk neutral density: proposed method vs Breeden-Litzenberger}
    \label{fig:Comparison of risk neutral density}
\end{minipage}
\end{figure}

Suppose the first option maturity to be considered is $T_1 = 2$. Given the previously generated IVs with random noise, Figure \ref{fig:combined-opt} presents the recovered IVs using local quadratic regression (LQR). Despite the presence of noise and potential arbitrage in the inputs, the LQR method effectively recovers the true IV with great accuracy. From IVs calibrated using local quadratic regression, we construct the lognormal mixture tails. Table \ref{tab:ssiv log mix param} shows the parameters obtained, where $K_L$ and $K_U$ represents the minimal and maximal observed strike prices. Figure \ref{fig:Lognormal Mixture Approximation under Pasting} shows the corresponding estimated risk neutral density with lognormal mixture tails.

\begin{figure}[ht]
\centering
\begin{minipage}{0.45\linewidth}
\includegraphics[width=\linewidth]{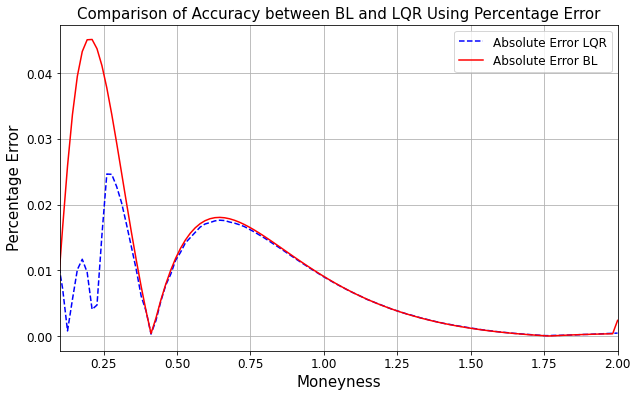}
    \caption{Estimation error for risk neutral density: proposed method vs Breeden-Litzenberger}
    \label{fig:Comparison of Accuracy between BL and LQR Using Absolute Error}
\end{minipage}
\hfill
\begin{minipage}{0.45\linewidth}
\includegraphics[width=\linewidth]{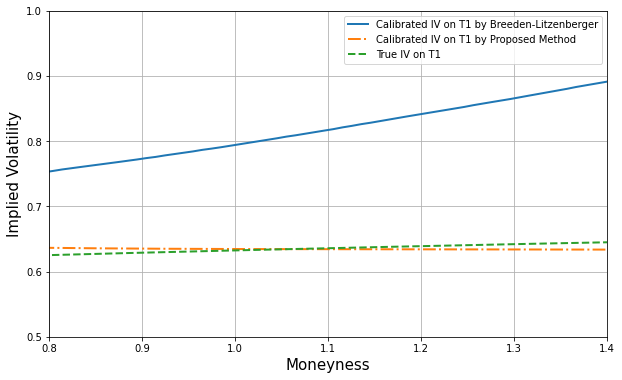}
    \caption{Calibrated IV: proposed method vs Breeden-Litzenberger}
    \label{fig:Comparison between Compound method and Breeden-Litzenberger}
\end{minipage}
\end{figure}

\begin{table}[ht]
\centering
\caption{Parameters for Lognormal Mixture Tails}
\label{tab:ssiv log mix param}
\begin{tabular}{lrr}
\toprule
Parameter &  Lower (i=L) &  Upper (i=U) \\
\midrule
  $\lambda_i$ & 0.879196378130754 & 0.021345797160246 \\
  $v_{i,1}$ & 0.999932326888895 & 0.138484172035767 \\
  $v_{i,2}$ & 0.696842788545524 & 0.946507489626882 \\
  $\mu_{i,1}$ & -0.035711890633837 & 0.326246597279264 \\
  $\mu_{i,2}$ & -0.876812461936515 & -0.507637416676997 \\
  $K_i$ & 6.016806722689075 & 159.8655462184874 \\
\bottomrule
\end{tabular}
\end{table}

To compare to the Breeden-Litzenberger approach, we consider a range of $[6,160]$ for the strike price.  
For the Breeden-Litzenberger approach, spline interpolation and extrapolation are used, 
with boundary conditions ensuring well-defined extrapolations. 
To optimize the performance of the Breeden-Litzenberger method, we use 120 evenly spaced strike prices ranging from 5 to 200, since the performance of the Breeden-Litzenberger method relies heavily on the accuracy of finite difference, which requires a larger number of observations.

Under these settings, we compare the two approaches in Figures \ref{fig:Comparison of risk neutral density}, \ref{fig:Comparison of Accuracy between BL and LQR Using Absolute Error}, and \ref{fig:Comparison between Compound method and Breeden-Litzenberger}. In Figure \ref{fig:Comparison of risk neutral density}, "LQR RND" shows the risk neutral density constructed using our proposed approach with local quadratic regression and lognormal mixture tails. "BL RND" represents the risk neutral density constructed using the Breeden-Litzenberger method. "True RND" represents the true risk neutral density in the Heston-like SSVI model. 

Although both "LQR RND" and "BL RND" seem to be close to the true risk neutral density, our proposed method provides a more accurate estimation. Figure \ref{fig:Comparison of Accuracy between BL and LQR Using Absolute Error} shows the absolute error for both estimated risk neutral densities. It can be seen that the proposed method (dashed line) achieves better accuracy compared to the Breeden-Litzenberger method (solid line). The latter shows notable errors for moneyness that is low. This leads to much larger errors for calibrated implied volatilities. In Figure \ref{fig:Comparison between Compound method and Breeden-Litzenberger}, we compare the implied volatilities (IVs) obtained from both methods to true IVs (represented by the dashed line). The solid curve, generated using the Breeden-Litzenberger method, deviates significantly from the true IV. In contrast, IVs calculated using our proposed method (represented by the dash-dotted line) closely align with the true IV.

\subsection{Numerical Experiment Based on TSLA Market Data}

\label{sec:tsla}
We conduct an experiment using the TSLA market data on July 1st, 2020. Since TSLA does not pay dividends, we treat the American call options as European call options and adjust the data to remove the effects of the interest rate term structure. The three maturities we select are 2020/09/18, 2020/10/16, and 2021/01/15, when sufficient market observations were available for call options. The option data was sourced from OptionMetrics.The implied volatilities used as calibration inputs 
are computed from mid quotes (midpoint of best bid and best ask) 
after standard filtering for liquidity and maturity.

We comment here that to handle American call option observables 
when the stock pays dividends or American put options in general, 
one can resort to an industry standard recipe: assuming dividends 
are fully proportional, one can invert an implied volatility from a 
given American option market observed price via an American option 
pricer (either a tree or a PDE engine). We then use the same implied 
volatility to feed into a European Black--Scholes pricer to obtain a 
European ``observable''. From these European prices, one can apply 
the usual Breeden--Litzenberger formula to obtain the risk neutral density.

Since the Bass construction assumes a zero interest rate, the data need to be regularized accordingly. 
Assuming the starting maturity is $T_0=0$, let $F(T)$ denote the option-implied forward price at maturity $T$.
For the TSLA case study, implied volatilities are obtained directly from 
OptionMetrics and are computed from mid quotes.
Since TSLA pays no dividends, call options carry no early-exercise premium, 
so the OptionMetrics implied volatilities for calls coincide with 
European implied volatilities.
For puts, we reprice at the OptionMetrics implied volatility using the 
Black--Scholes formula to obtain European-equivalent put prices.  
The forward price is then recovered from European put--call parity,
\begin{equation}\label{eq:fwd-pcp}
F(T)=K+\frac{C^{Eu}(K,T)-P^{Eu}(K,T)}{DF(T)}.
\end{equation}
In practice, we evaluate~\eqref{eq:fwd-pcp} across a range of 
near-the-money strikes where both calls and puts are liquid 
and take the median to obtain a robust estimate of $F(T)$.
We define the moneyness at maturity $T$ as $km(T)=\frac{K(T)}{F(T)}$ where $K(T)$ is the real strike at maturity $T$. As such, the normalized call option price is calculated as:
$$C_N(T,km(T))=\frac{C(T,K(T))}{F(T)}.$$

In implementation,
the discount factor $DF(T)$ used in~\eqref{eq:fwd-pcp} and for present-value 
discounting is computed as $DF(T)=e^{-r(T)\,T}$, where $r(T)$ is the 
continuously compounded zero rate at maturity $T$ obtained from the 
OptionMetrics zero curve, with standard interpolation across maturities 
when needed.

\begin{figure}[ht]
\centering
\begin{minipage}{0.53\linewidth}
\includegraphics[width=\linewidth]{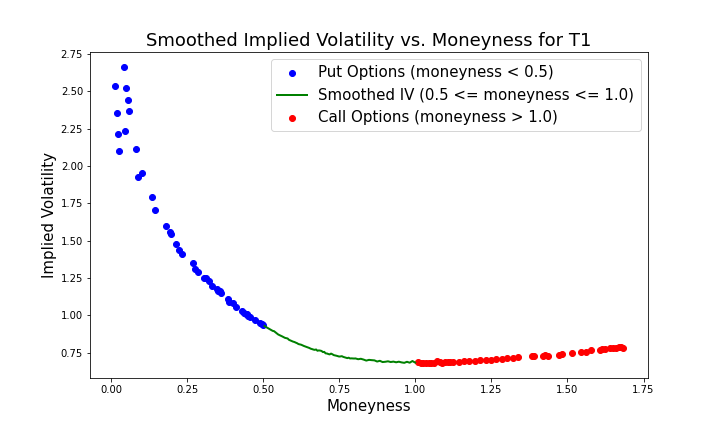}
    \caption{Smoothed IV for maturity T1}
    \label{fig:Smoothed IV on Maturity T1}
\end{minipage}
\hfill
\begin{minipage}{0.45\linewidth}
\includegraphics[width=\linewidth]{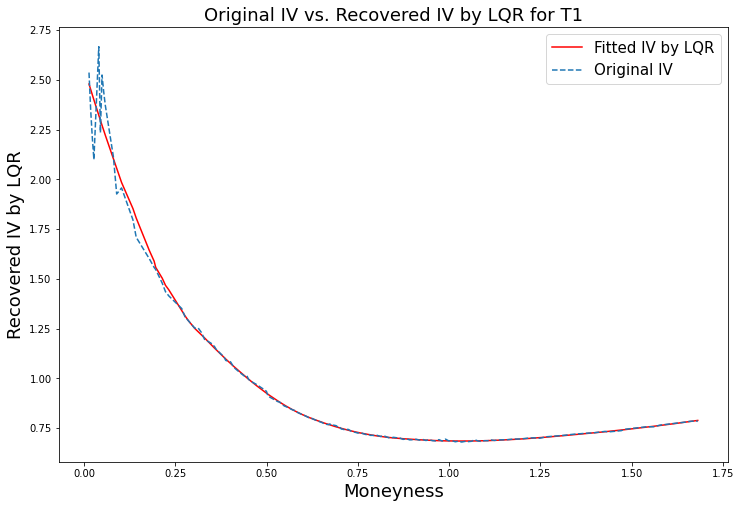}
    \caption{LQR fitted IV for T1}
    \label{fig:LQR Fitting for Maturity T1}
\end{minipage}
\end{figure}

To avoid potential artificial jumps in implied volatility (IV) at the at-the-money (ATM) region, we applied a smoothing procedure to the IV curves. We use a blending approach as described in \citep{birru2010impact} and \citep{alexiou2023pricing}. In this approach, the IVs of put and call options with strike prices within a specified range close to the underlying spot price $S_0$ are blended as follows:
\begin{align*}
    \hat{IV}(K)=w\cdot IV_{put}(K)+(1-w)\cdot IV_{call}(k),
\end{align*}
with $w=\frac{K_{max}-K}{K_{max}-K_{min}}$, and $K_{max}$ and $K_{min}$ are the maximum and minimum strike prices of the range, respectively. For our experiment, we focus on calibrating the market smile on the call option side. We therefore choose the blending region as $(0.5S_0,S_0)$. For illustration, result of smoothing for maturity T1 is shown in Figure \ref{fig:Smoothed IV on Maturity T1}.

\begin{figure}[ht]
\centering
\begin{minipage}{0.45\linewidth}
\includegraphics[width=\linewidth]{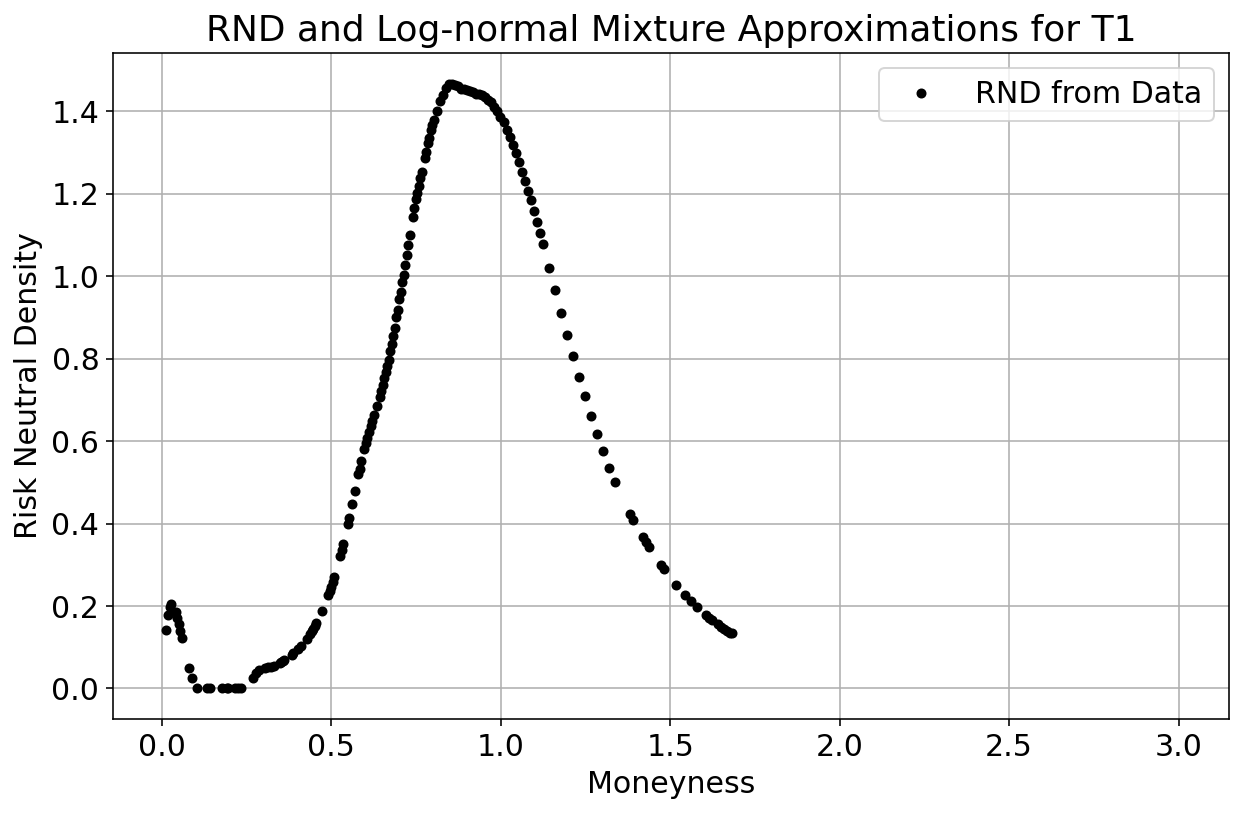}
    \caption{Risk neutral density for maturity T1 from LQR fitted implied volatility}
    \label{fig:Mixture Log-normal approximation for Maturity T1--Whole curve}
\end{minipage}
\hfill
\begin{minipage}{0.45\linewidth}
\includegraphics[width=\linewidth]{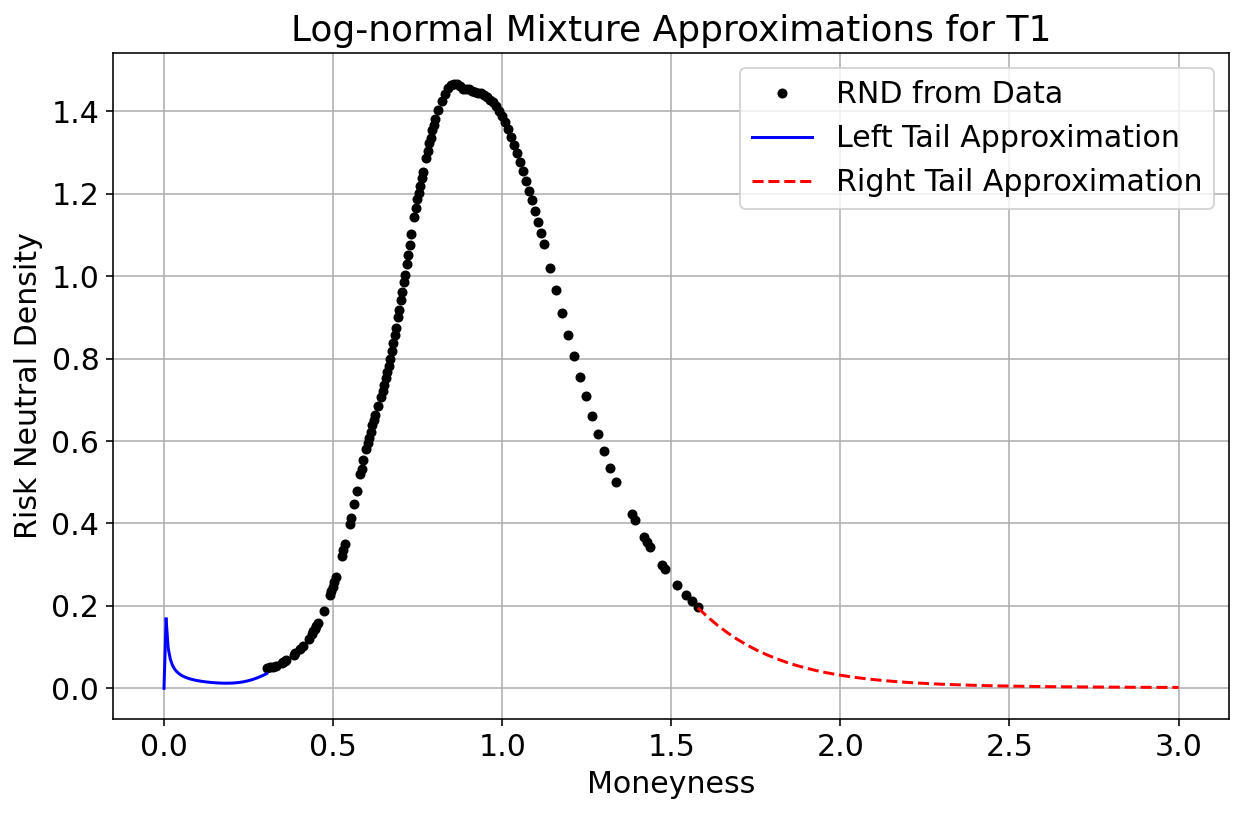}
    \caption{Lognormal mixture tails of the risk neutral density for maturity T1}
    \label{fig:Mixture Log-normal approximation for Maturity T1}
\end{minipage}
\end{figure}

After cleaning and processing the data, we apply the LQR model to the three maturities to generate arbitrage-free IV curves and the corresponding values of the risk neutral densities. The fitted IV curve for maturity T1 is shown in Figure \ref{fig:LQR Fitting for Maturity T1}. The risk neutral density calculated based on this IV curve can be seen in Figure \ref{fig:Mixture Log-normal approximation for Maturity T1--Whole curve}.

Then we complete the risk neutral densities with lognormal mixture tails, as described in Section 3.2. Note that the density in Figure \ref{fig:Mixture Log-normal approximation for Maturity T1--Whole curve} is bimodal, with a minor mode near zero. This could lead to numerical difficulties in solving the nonlinear system in Section 3.2. To address this, we slightly reduce the domain $[K_L,K_U]$ to exclude this region, and impose constraints $v_{L,1} \geq 1$, $v_{L,2} \in (0,1)$, and $\lambda_L > 0.5$ to create a desired lognormal mixture tail on the left with a minor mode. Table \ref{tab:Market lognormal param} presents the parameters of the lognormal mixture tails for the risk neutral densities corresponding to all the three maturities. Figure \ref{fig:Mixture Log-normal approximation for Maturity T1} shows the risk neutral density completed with lognormal mixture tails for maturity T1. The thick black curve is computed from LQR fitted IVs, and the blue solid left tail and the red dashed right tail are lognormal mixtures with parameters given in the T1 column of Table \ref{tab:Market lognormal param}.

\begin{table}[ht]
\centering
\caption{Parameter settings for lognormal mixture tails}
\label{tab:Market lognormal param}
\begin{tabular}{cccc}
\toprule
Parameters & T1 & T2 & T3 \\
\midrule
$\lambda_L$ & 0.891814729 & 0.730072506 & 0.999997865 \\
$v_{L1}$ & 11.83540063 & 35.27086006 & 5.874220433 \\
$v_{L2}$ & 0.389034015 & 0.442752001 & 0.117462989 \\
$\mu_{L1}$ & 25.61363809 & 75.02042380 & 9.323137361 \\
$\mu_{L2}$ & -0.312016942 & -0.153184102 & -1.246629261 \\
$\lambda_U$ & 0.818069793 & 0.927681918 & 0.971590088 \\
$v_{U1}$ & 0.283361639 & 0.345919720 & 0.437880299 \\
$v_{U2}$ & 1.167721500 & 1.708909349 & 2.078893403 \\
$\mu_{U1}$ & -0.013315716 & -0.065021265 & -0.091679922 \\
$\mu_{U2}$ & -1.408634421 & -2.074760445 & -1.947687673 \\
$L_{\text{cdf}}$ & 0.011757319 & 0.015447680 & 0.033174055 \\
$U_{\text{cdf}}$ & 0.957228077 & 0.929827574 & 0.870975267 \\
\bottomrule
\end{tabular}
\end{table}

After we obtain arbitrage-free risk neutral densities for all maturities, we use spline interpolation and extrapolation and compute the probability distribution functions and their inverse corresponding to the densities. These are required for the Bass construction. 

\begin{figure}[ht]
\centering
\begin{minipage}{0.45\linewidth}
\includegraphics[width=\linewidth]{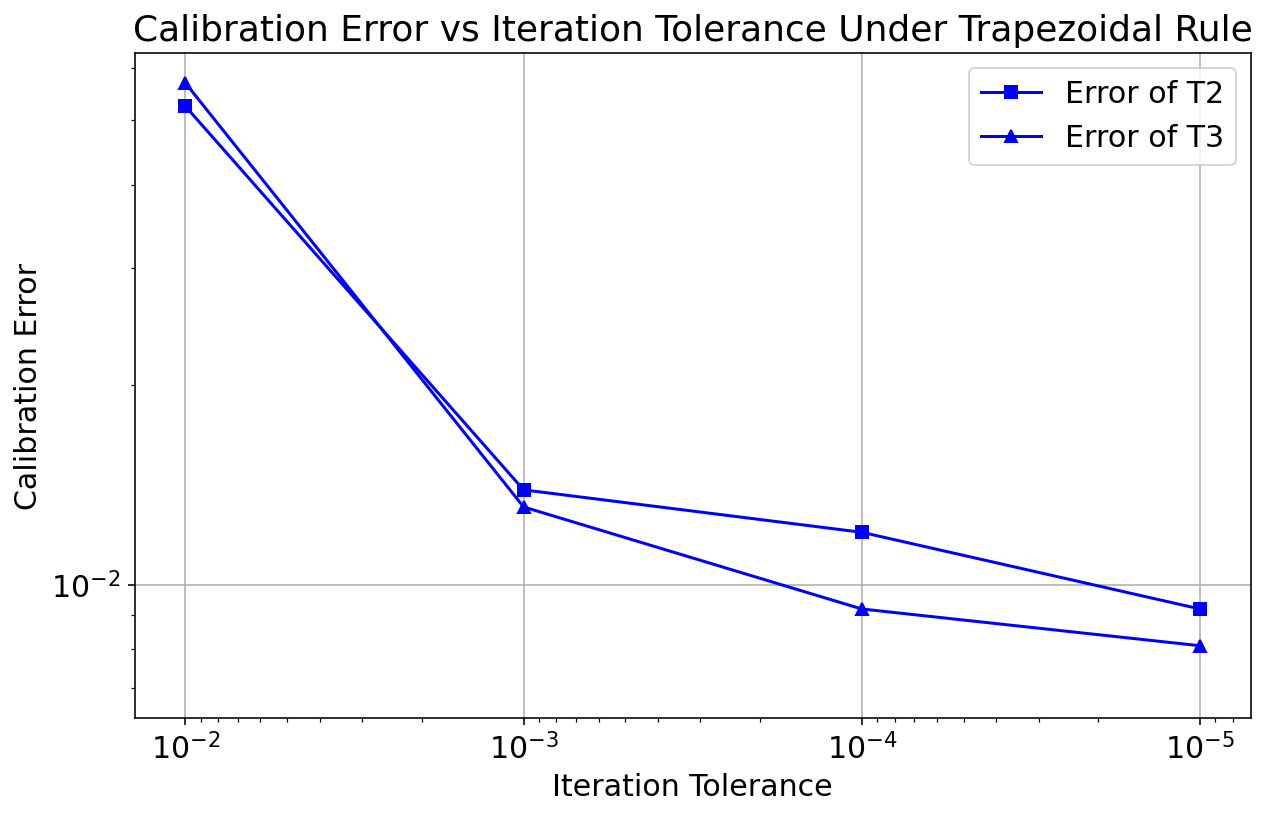}
    \caption{Calibration error vs. iteration error tolerance for the TSLA case}
    \label{fig:Calibration Error VS Iteration Tolerance under Trapezoidal Rule Scheme of Market Case}
\end{minipage}
\hfill
\begin{minipage}{0.45\linewidth}
\includegraphics[width=\linewidth]{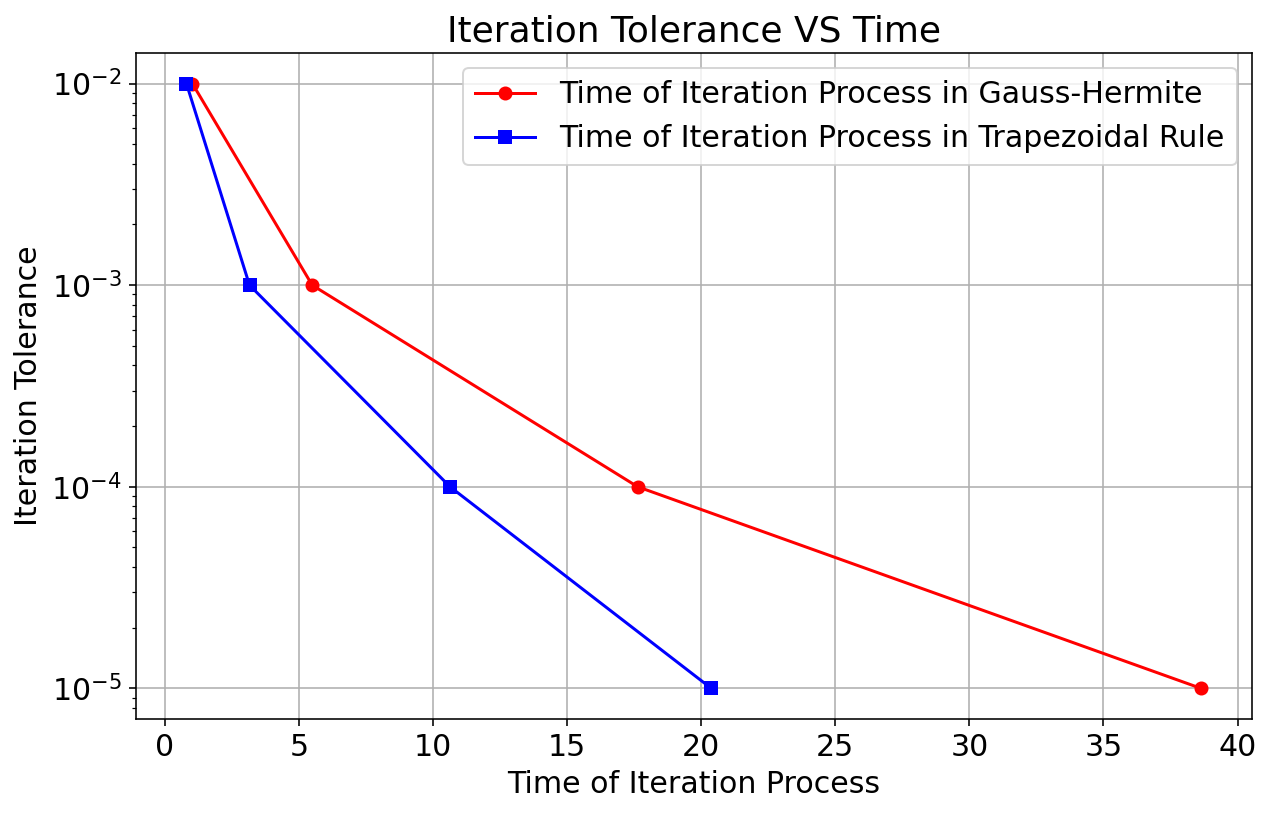}
    \caption{Time for solving fixed point problems in TSLA case: trapezoidal rule vs. Gauss Hermite quadrature}
    \label{fig:Iteration Time for Two Numerical Scheme for Market Data}
\end{minipage}
\end{figure}

Figures \ref{fig:Calibration Error VS Iteration Tolerance under Trapezoidal Rule Scheme of Market Case} depicts the relationship between calibration accuracy and iteration error tolerance for maturities T2 and T3. Detailed numerical results are provided in table \ref{tab:Results from Gauss Hermite Quadrature} and \ref{tab:Results from Trapezoidal Rule Scheme}. Option prices and the corresponding implied volatilities in the calibrated model are computed using Monte Carlo simulation with a sample size of around $7 \times 10^{6}$. Note that no fixed point problem needs to be solved for options pricing and calculating the IVs for the first maturity. By fixing the simulation random seeds, we obtain a constant calibration error, determined solely by the Monte Carlo sample size, for this maturity. Figure \ref{fig:Calibration Error VS Iteration Tolerance under Trapezoidal Rule Scheme of Market Case} and Tables \ref{tab:Results from Gauss Hermite Quadrature}, \ref{tab:Results from Trapezoidal Rule Scheme} show that reducing the iteration error tolerance from $10^{-2}$ to $10^{-3}$ significantly reduces the calibration error. However, reducing the tolerance further does not lead to further significant improvement, since the calibration error will eventually be dominated by the error due to Monte Carlo sample size. 

Figure \ref{fig:Iteration Time for Two Numerical Scheme for Market Data} compares the computational time needed for solving the fixed point problems using the trapezoidal rule and Gauss-Hermite quadrature. It shows that, when higher calibration accuracy is desired and hence smaller iteration error control level is used, the trapezoidal rule will be much faster and is hence preferred.

\begin{figure}
\centering
\begin{minipage}{0.4\linewidth}
\includegraphics[width=\linewidth]{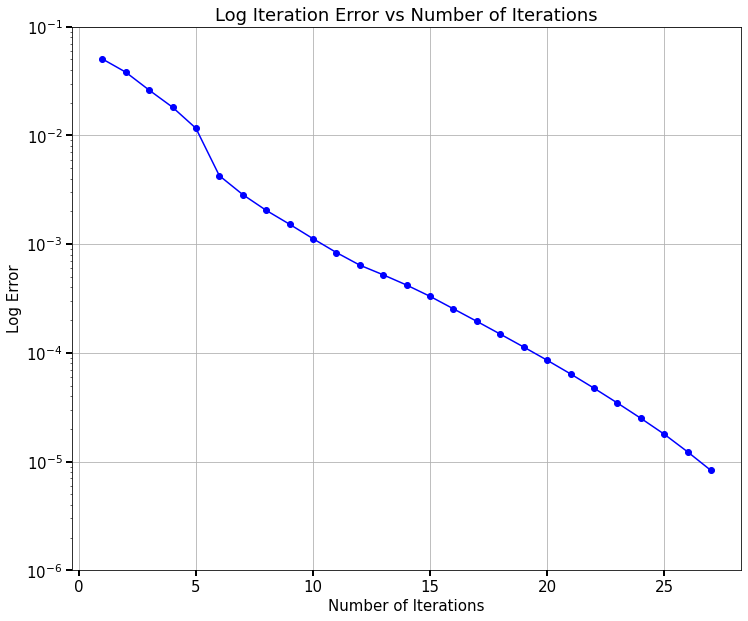}
    \caption{Log iteration error vs. number of iterations in TSLA case}
    \label{fig:Iteration Tolerance VS Iteration number under Market Case}

\end{minipage}
\hfill
\begin{minipage}{0.55\linewidth}
\includegraphics[width=\linewidth]{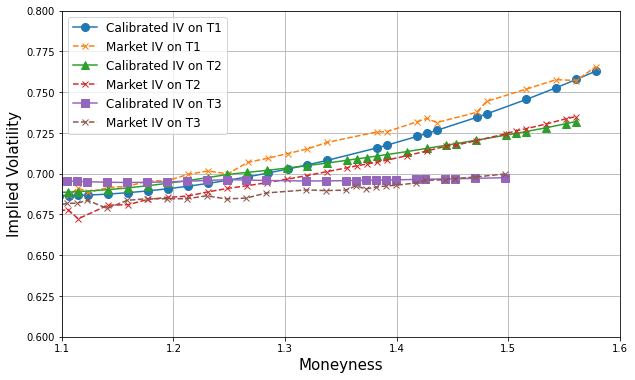}
    \caption{Observed IVs vs. IVs computed from option prices in the calibrated Bass-LV model: iteration error tolerance $=10^{-4}$}
    \label{fig:IV Curve Fitting under 1e-4 Iteration Tolerance}

\end{minipage}
\end{figure}

Figure \ref{fig:Iteration Tolerance VS Iteration number under Market Case} demonstrates the linear convergence of the fixed-point algorithm, as establised in \citep{acciaio2023calibration}, in the TSLA case study. Finally, Figure \ref{fig:IV Curve Fitting under 1e-4 Iteration Tolerance} shows the implied volatilities of options prices calculated in the calibrated Bass-LV model. The iteration error tolerance is set at $10^{-4}$. The Figure shows the effectiveness of the proposed methods in fitting even rather atypical market implied volatilities. 

\begin{table}[H]
\centering
\caption{Time and calibration error in the TSLA case: Gauss-Hermite quadrature}
\label{tab:Results from Gauss Hermite Quadrature}
\scalebox{1}{
\begin{tabular}{ccccc}
\toprule
Iteration  & Iteration  & Calibration   & Calibration   & Calibration \\
tolerance  & time (seconds)       & error ($T_1$) & error ($T_2$) & error ($T_3$) \\
\midrule
1.0e-2 & 1.0 & 7.0e-3 & 4.8e-2 & 5.7e-2 \\
1.0e-3 & 5.5  & 7.0e-3 & 1.7e-2 & 1.9e-2 \\
1.0e-4 & 17.6 & 7.0e-3 & 8.5e-3 & 8.3e-3 \\
1.0e-5 & 38.6 & 7.0e-3 & 8.4e-3 & 7.8e-3 \\
\bottomrule
\end{tabular}}
\end{table}

\begin{table}[H]
\centering
\caption{Time and calibration error in the TSLA case: trapezoidal rule}
\label{tab:Results from Trapezoidal Rule Scheme}
\scalebox{1}{
\begin{tabular}{ccccc}
\toprule
Iteration  & Iteration  & Calibration   & Calibration   & Calibration \\
tolerance  & time (seconds)       & error ($T_1$) & error ($T_2$) & error ($T_3$) \\
\midrule
1.0e-2 & 0.8 & 7.0e-3 & 5.3e-2 & 5.0e-2 \\
1.0e-3 & 3.2 & 7.0e-3 & 1.4e-2 & 1.3e-2 \\
1.0e-4 & 10.6 & 7.0e-3 & 1.2e-2 & 9.2e-3 \\
1.0e-5 & 20.4 & 7.0e-3 & 9.2e-3 & 8.1e-3 \\
\bottomrule
\end{tabular}}
\end{table}

\section{Conclusions}

This paper studies the practical implementation of the Bass--LV calibration
pipeline, whose calibrate-once, query-anywhere property makes reliable
marginal inputs and efficient repeated convolutions essential.
On the marginal construction side, we unify adaptive local quadratic
estimation in the liquid region with lognormal-mixture tail completion
into a coherent pipeline.
At the middle--tail stitching points, we enforce boundary, monotonicity,
and calendar-consistency constraints. The matching order of these
constraints is chosen to be compatible with the convergence requirements
of the downstream convolution scheme.
On the computational side, we provide an optimal parameter selection and
convergence analysis. It shows that trapezoidal-rule based convolutions are
both more accurate and faster than commonly used Gauss--Hermite quadrature
under the limited smoothness typical of fitted marginals.
Numerical experiments in standard models and in a detailed market case
study demonstrate that the proposed marginal construction improves
calibration robustness and accuracy, while the trapezoidal convolution
scheme accelerates the fixed-point iterations.
The two contributions are co-designed: the convergence analysis dictates the regularity
requirements on the marginal inputs, and the construction pipeline delivers them,
making the overall Bass--LV calibration both robust and fast.

\section{Appendix - Proofs}

\subsection{Proof of Lemma~\ref{lem:bass_time_interp}}
\label{app:time_interp}

Let $(S_t)_{t\in[T_1,T_2]}$ be the Bass--LV martingale.

\paragraph{Butterfly-arbitrage-freeness.}
Fix $t\in(T_1,T_2)$.
The crucial distinction of the Bass--LV interpolation from standard interpolation schemes is that
the Bass--LV construction produces a genuine martingale $(S_t)$,
so the intermediate call price is not obtained by interpolating
implied volatilities or total variances and then inverting the
Black--Scholes formula.  Instead, it is \emph{defined directly}
by the marginal law $\mu_t = \mathrm{Law}(S_t)$:
\[
C(t,K) = E\big[(S_t - K)^+\big]
       = \int_0^\infty (s-K)^+\,\mu_t(ds).
\]
We now show that this representation automatically ensures
that $K\mapsto C(t,K)$ is nonincreasing and convex.

\noindent \emph{Monotonicity.}\;
For $K_1 < K_2$ and every $s \ge 0$,
$(s - K_1)^+ \ge (s - K_2)^+$ pointwise.
Integrating against the non-negative measure $\mu_t$ preserves
the inequality:
\[
C(t,K_1)
= \int_0^\infty (s-K_1)^+\,\mu_t(ds)
\;\ge\;
\int_0^\infty (s-K_2)^+\,\mu_t(ds)
= C(t,K_2).
\]

\noindent \emph{Convexity.}\;
For any $\lambda \in [0,1]$, $K_1, K_2 > 0$, and every fixed
$s \ge 0$, the map $K \mapsto (s-K)^+ = \max\{s-K,\,0\}$ is convex, so
\begin{equation}\label{eq:ptwise_convex}
\bigl(s - (\lambda K_1 + (1-\lambda)K_2)\bigr)^+
\;\le\;
\lambda\,(s-K_1)^+ + (1-\lambda)\,(s-K_2)^+.
\end{equation}
Integrating both sides of~\eqref{eq:ptwise_convex} against $\mu_t$
yields
\[
C\bigl(t,\,\lambda K_1 + (1-\lambda)K_2\bigr)
\;\le\;
\lambda\,C(t,K_1) + (1-\lambda)\,C(t,K_2).
\]
In contrast, standard time-interpolation methods
(e.g.\ linear blending of total implied variance as
in~\eqref{eq:total_var_interp}) operate on derived quantities
$\omega(t,k)$ and recover call prices only after applying the
Black--Scholes inversion $\omega \mapsto C$.  Since this inversion
is nonlinear, convexity of $K \mapsto C(t,K)$ can be lost even
when both boundary slices $\omega(T_1,\cdot)$ and
$\omega(T_2,\cdot)$ individually satisfy it.
The Bass--LV construction avoids this entirely:
because $(S_t)$ is a martingale, every intermediate call price
is an expectation of a convex payoff under a probability measure,
and convexity in $K$ is inherited automatically.

\paragraph{Calendar-arbitrage-freeness.}
Fix $K>0$ and let $T_1<s<t<T_2$.
Since $(S_u)$ is a martingale, $S_s=E[S_t\mid\mathcal{F}_s]$.
The payoff $\phi_K(x):=(x-K)^+$ is convex, so by conditional Jensen inequality,
\[
\phi_K(S_s)\le E\big[\phi_K(S_t)\mid\mathcal{F}_s\big].
\]
Taking expectations yields $C(s,K)\le C(t,K)$, hence $t\mapsto C(t,K)$ is nondecreasing.
\qed

\subsection{Proof of Propositions \ref{proposition:wing_option_price} and \ref{proposition:wing_iv_asymptotics}}
\label{appendix:proof_wing}

\begin{proof}[\textbf{Proof of Proposition~\ref{proposition:wing_option_price}}]
Consider a lognormal r.v. $X$ with parameters $(\eta,v^2)$. Define $\mu := \ln(\eta) - \frac{1}{2}v^2$ and let $X = \exp(\mu + v Z)$ with $Z \sim N(0,1)$, so that
\[
\mathbb{E}[X] = \exp\!\Big(\mu + \tfrac{1}{2}v^2\Big) = \eta.
\]
For any strike $K > 0$, the expected call payoff decomposes as
\begin{equation}
\mathbb{E}\big[(X - K)^+\big] = \mathbb{E}\big[X\,\mathbf{1}_{\{X > K\}}\big] - K\,\mathbb{P}(X > K).
\label{eq:call_decomp}
\end{equation}
Since $X > K$ if and only if $\mu + v Z > \ln K$, or equivalently, $Z > c$ where
\[
c := \frac{\ln K - \mu}{v},
\]
we obtain
\[
\mathbb{P}(X > K) = \mathbb{P}(Z > c) = 1 - \Phi(c) = \Phi(-c).
\]
For the first term, let $\phi(\cdot)$ denote the standard normal density,
\[
\mathbb{E}\big[X\,\mathbf{1}_{\{X > K\}}\big] = \mathbb{E}\big[e^{\mu + v Z}\,\mathbf{1}_{\{Z > c\}}\big] = e^{\mu}\int_c^{\infty} e^{v z}\,\phi(z)\,dz
\]
\[
= \exp\!\Big(\mu+\tfrac{1}{2}v^2\Big)\int_c^{\infty}\phi(z - v)\,dz = \exp\!\Big(\mu+\tfrac{1}{2}v^2\Big)\,\Phi(v - c)
= \eta\,\Phi(v - c).
\]
Define
\[
d_2(K) := \frac{\ln(\eta/K) - \frac{1}{2}v^2}{v}, \qquad d_1(K) := d_2(K) + v = \frac{\ln(\eta/K) + \frac{1}{2}v^2}{v}.
\]
Observe that $c = -d_2(K)$ and $v - c = d_1(K)$. Substituting into~\eqref{eq:call_decomp} gives
\begin{equation}
\mathbb{E}\big[(X - K)^+\big] = \eta\,\Phi\!\big(d_1(K)\big) - K\,\Phi\!\big(d_2(K)\big),
\label{eq:single_call_result}
\end{equation}
which is the Black--Scholes forward call formula with forward $\eta$ and total log-volatility $v$. This establishes~\eqref{eq:lognormal_call}.

For the put payoff, we write
\[
\mathbb{E}\big[(K - X)^+\big] = K\,\mathbb{P}(X < K) - \mathbb{E}\big[X\,\mathbf{1}_{\{X < K\}}\big].
\]
Since $\mathbb{P}(X < K) = \Phi(c) = \Phi(-d_2(K))$, and
\[
\mathbb{E}\big[X\,\mathbf{1}_{\{X < K\}}\big] = e^{\mu}\int_{-\infty}^{c} e^{v z}\,\phi(z)\,dz = \eta\,\Phi(c - v) = \eta\,\Phi(-d_1(K)),
\]
where $c - v = -(d_2(K) + v) = -d_1(K)$, we obtain
\begin{equation}
\mathbb{E}\big[(K - X)^+\big] = K\,\Phi\!\big({-d_2(K)}\big) - \eta\,\Phi\!\big({-d_1(K)}\big),
\label{eq:single_put_result}
\end{equation}
establishing~\eqref{eq:lognormal_put}.

For the upper tail density $q^U(x;\theta_U) = \lambda_U\,\ell(x;\eta_{U,1},v_{U,1}^2) + (1-\lambda_U)\,\ell(x;\eta_{U,2},v_{U,2}^2)$, for $K>K_U$, linearity of integration and~\eqref{eq:single_call_result} yield
\[
\int_K^{\infty}(x - K)\,q^U(x;\theta_U)\,dx = \lambda_U\,C^{(1)}(K) + (1 - \lambda_U)\,C^{(2)}(K),
\]
where for $j = 1, 2$,
\[
C^{(j)}(K) = \eta_{U,j}\,\Phi\!\big(d_1(K;\eta_{U,j},v_{U,j})\big) - K\,\Phi\!\big(d_2(K;\eta_{U,j},v_{U,j})\big).
\]
The result for puts with $K<K_L$ follows in the similar way. This completes the proof of Proposition~\ref{proposition:wing_option_price}.
\end{proof}

\begin{proof}[\textbf{Proof of Proposition~\ref{proposition:wing_iv_asymptotics}}]
We establish the right-wing result~\eqref{eq:right_wing_iv} in three parts.
The left-wing result~\eqref{eq:left_wing_iv} is then obtained by an analogous argument applied to put prices.

\medskip
\noindent\textbf{Part 1: Logarithmic decay rate of a single component.}
Fix $j\in\{1,2\}$ and write $v:=v_{U,j}$, $\eta:=\eta_{U,j}$.
The component call price from Proposition~\ref{proposition:wing_option_price} is
\[
C^{(j)}(K)=\eta\,\Phi\!\big(d_1(K)\big)-K\,\Phi\!\big(d_2(K)\big),
\]
with $d_2(K)=\bigl(\ln(\eta/K)-v^2/2\bigr)/v$ and $d_1(K)=d_2(K)+v$.
As $K\to\infty$, both $d_1(K)$ and $d_2(K)$ tend to $-\infty$ with
\begin{equation}\label{eq:d2_leading}
d_2(K)=-\frac{\ln K}{v}+O(1),\qquad d_1(K)=-\frac{\ln K}{v}+O(1).
\end{equation}
Applying the standard Mills-ratio asymptotic
$\Phi(-t)\sim\phi(t)/t$ as $t\to+\infty$
with $t=|d_2(K)|$ gives
\[
\Phi\!\big(d_2(K)\big)=\Phi\!\big(-|d_2(K)|\big)
\sim\frac{\phi\!\big(|d_2(K)|\big)}{|d_2(K)|}
=\frac{1}{|d_2(K)|\sqrt{2\pi}}\exp\!\Bigl(-\tfrac{d_2(K)^2}{2}\Bigr).
\]
Since $d_2(K)=\bigl(\ln(\eta/K)-v^2/2\bigr)/v$ and $d_1(K)=d_2(K)+v$,
both tend to $-\infty$ as $K\to\infty$.
Write $|d_i|:=-d_i>0$ for $i=1,2$.
Applying the Mills-ratio asymptotic
$\Phi(-t)=\frac{\phi(t)}{t}\bigl(1+O(t^{-2})\bigr)$ as $t\to+\infty$
to $\Phi(d_i)=\Phi(-|d_i|)$ gives
\begin{equation}\label{eq:mills_applied}
C^{(j)}(K)
=\frac{\eta\,\phi(|d_1|)}{|d_1|}\bigl(1+O(|d_1|^{-2})\bigr)
-\frac{K\,\phi(|d_2|)}{|d_2|}\bigl(1+O(|d_2|^{-2})\bigr).
\end{equation}

From $d_1=d_2+v$, we have $d_1^2=d_2^2+2v\,d_2+v^2$, hence
\[
\frac{\eta\,\phi(|d_1|)}{\phi(|d_2|)}
=\eta\,\exp\!\Bigl(-\tfrac{d_1^2-d_2^2}{2}\Bigr)
=\eta\,\exp\!\bigl(-v\,d_2-\tfrac{v^2}{2}\bigr).
\]
Substituting the definition $v\,d_2=\ln(\eta/K)-v^2/2$ yields
$\exp(-v\,d_2-v^2/2)=\exp\bigl(-\ln(\eta/K)\bigr)=K/\eta$,
and therefore
\begin{equation}\label{eq:eta_phi_identity}
\eta\,\phi(|d_1|)=K\,\phi(|d_2|).
\end{equation}
Applying~\eqref{eq:eta_phi_identity} to rewrite the first term
in~\eqref{eq:mills_applied}, we factor out $K\,\phi(|d_2|)$:
\[
C^{(j)}(K)
=K\,\phi(|d_2|)\left[
\frac{1}{|d_1|}\bigl(1+O(|d_1|^{-2})\bigr)
-\frac{1}{|d_2|}\bigl(1+O(|d_2|^{-2})\bigr)
\right].
\]
The leading part of the bracket is
\[
\frac{1}{|d_1|}-\frac{1}{|d_2|}
=\frac{|d_2|-|d_1|}{|d_1|\,|d_2|}.
\]
Since $|d_1|=|d_2|-v$ (both $d_1$ and $d_2$ are negative and $d_1=d_2+v$),
we have $|d_2|-|d_1|=v$, giving
\[
\frac{1}{|d_1|}-\frac{1}{|d_2|}=\frac{v}{|d_1|\,|d_2|}.
\]
The Mills-ratio remainders contribute terms of order
$O(|d_i|^{-3})$ to the bracket, whereas the leading term is
$O(|d_i|^{-2})$.
Since $|d_2|=(\ln K)/v+O(1)$, the relative error is
$O(|d_2|^{-1})=O(1/\ln K)$, and we obtain
\begin{equation}\label{eq:Cj_asymp}
C^{(j)}(K)
=K\,\phi\!\big(|d_2(K)|\big)\,\frac{v}{|d_1(K)|\,|d_2(K)|}
\bigl(1+O(1/\ln K)\bigr).
\end{equation}

Taking logarithms, with similar analysis, the leading-order behavior of the exponent is
\[
-\ln C^{(j)}(K)
=\frac{(\ln K)^2}{2\,v^2}+O(\ln K),
\]
which yields the key relation
\begin{equation}\label{eq:log_decay_single}
\lim_{K\to\infty}\frac{-\ln C^{(j)}(K)}{(\ln K)^2}=\frac{1}{2\,v_{U,j}^2}.
\end{equation}

\medskip
\noindent\textbf{Part 2: Dominant component of the mixture.}
The mixture call price is
$C_{\mathrm{mix}}(K)=\lambda_U\,C^{(1)}(K)+(1-\lambda_U)\,C^{(2)}(K)$.
Without loss of generality, suppose $v_{U,1}\ge v_{U,2}$, so that $v_U^\ast=v_{U,1}$.

If $v_{U,1}>v_{U,2}$, consider the ratio $C^{(2)}(K)/C^{(1)}(K)$.
From~\eqref{eq:log_decay_single},
\[
\lim_{K\to\infty}\frac{-\ln\bigl(C^{(2)}(K)/C^{(1)}(K)\bigr)}{(\ln K)^2}
=\frac{1}{2v_{U,2}^2}-\frac{1}{2v_{U,1}^2}>0,
\]
where the positivity follows from $v_{U,1}>v_{U,2}$.
Since $(\ln K)^2\to\infty$, the exponent $-\ln\bigl(C^{(2)}/C^{(1)}\bigr)\to+\infty$,
and hence $C^{(2)}(K)/C^{(1)}(K)\to 0$ as $K\to\infty$.
Since $$
C_{\operatorname{mix}}(K)=\lambda_U C^{(1)}(K)+\left(1-\lambda_U\right) C^{(2)}(K)=\lambda_U C^{(1)}(K)\left(1+\frac{\left(1-\lambda_U\right) C^{(2)}(K)}{\lambda_U C^{(1)}(K)}\right),
$$ we have
\[
C_{\mathrm{mix}}(K)=\lambda_U\,C^{(1)}(K)\bigl(1+o(1)\bigr),
\]
and since $-\ln\lambda_U$ is $O(1)$, it follows that
\begin{equation}\label{eq:log_decay_mix}
\lim_{K\to\infty}\frac{-\ln C_{\mathrm{mix}}(K)}{(\ln K)^2}
=\frac{1}{2\,(v_U^\ast)^2}.
\end{equation}
If $v_{U,1}=v_{U,2}=v_U^\ast$, both components share the same decay rate and~\eqref{eq:log_decay_mix} holds trivially.

\medskip
\noindent\textbf{Part 3: Extraction of the limiting implied volatility.}
Denote the Black--Scholes call price with forward $F>0$ and annualized
implied volatility $\sigma>0$ by
\[
C_{\mathrm{BS}}(K;\sigma):=F\,\Phi\!\bigl(\hat d_1(K;\sigma)\bigr)-K\,\Phi\!\bigl(\hat d_2(K;\sigma)\bigr),
\]
where $\hat d_{2}(K;\sigma):=\bigl(\ln(F/K)-\tfrac12\sigma^2\tau\bigr)/(\sigma\sqrt\tau)$ and $\hat d_1=\hat d_2+\sigma\sqrt\tau$.
Repeating the calculation in Part~1 with total volatility $\sigma\sqrt\tau$ in place of $v$, we obtain
\begin{equation}\label{eq:log_decay_bs}
\lim_{K\to\infty}\frac{-\ln C_{\mathrm{BS}}(K;\sigma)}{(\ln K)^2}=\frac{1}{2\,\sigma^2\tau}.
\end{equation}
The implied volatility $\sigma_{\mathrm{imp}}(K,\tau)$ is defined by
$C_{\mathrm{mix}}(K)=C_{\mathrm{BS}}\!\bigl(K;\sigma_{\mathrm{imp}}(K,\tau)\bigr)$.
Because $K\mapsto C_{\mathrm{BS}}(K;\sigma)$ is strictly decreasing in $K$ for each fixed $\sigma$,
and strictly increasing in $\sigma$ for each fixed $K$, the map $\sigma_{\mathrm{imp}}(K,\tau)$ is well-defined.

We now show $\sigma_{\mathrm{imp}}(K,\tau)\to v_U^\ast/\sqrt\tau$ by a sandwich argument.
Fix an arbitrary $\delta>0$ and set $\sigma^{\pm}:=(v_U^\ast\pm\delta)/\sqrt\tau$.
From~\eqref{eq:log_decay_bs},
\[
\lim_{K\to\infty}\frac{-\ln C_{\mathrm{BS}}(K;\sigma^-)}{(\ln K)^2}
=\frac{1}{2(v_U^\ast-\delta)^2}
>\frac{1}{2(v_U^\ast)^2}
=\lim_{K\to\infty}\frac{-\ln C_{\mathrm{mix}}(K)}{(\ln K)^2},
\]
so for all sufficiently large $K$,
$C_{\mathrm{mix}}(K)>C_{\mathrm{BS}}(K;\sigma^-)$,
which by monotonicity of $C_{\mathrm{BS}}$ in $\sigma$ implies
$\sigma_{\mathrm{imp}}(K,\tau)>\sigma^-=(v_U^\ast-\delta)/\sqrt\tau$.
An analogous argument using $\sigma^+$ gives
$\sigma_{\mathrm{imp}}(K,\tau)<(v_U^\ast+\delta)/\sqrt\tau$
for all sufficiently large $K$.
Since $\delta>0$ was arbitrary,
\[
\sigma_{\mathrm{imp}}(K,\tau)\;\longrightarrow\;\frac{v_U^\ast}{\sqrt\tau}
\qquad\text{as }K\to\infty.
\]

\medskip
\noindent\textbf{Left wing.}
For the left wing ($K\to 0$), define put prices
$P(K):=\int_0^K(K-x)\,q(x)\,dx$.
The lower tail density $q^L(\cdot;\theta_L)$ is a mixture of two lognormals
with log-variances $v_{L,1}^2$ and $v_{L,2}^2$.
Applying the put-price formula~\eqref{eq:lognormal_put} from
Proposition~\ref{proposition:wing_option_price} and repeating
Parts~1--3 with $K\to 0$ (equivalently $|\ln K|\to\infty$) and
total volatility $v_{L,j}$ in place of $v_{U,j}$, one obtains
\[
\lim_{K\to 0}\frac{-\ln P_{\mathrm{mix}}(K)}{(\ln K)^2}=\frac{1}{2(v_L^\ast)^2},
\]
where $v_L^\ast=\max\{v_{L,1},v_{L,2}\}$.
The Black--Scholes put price with implied volatility $\sigma$ satisfies the same
logarithmic decay rate $1/(2\sigma^2\tau)$, and the sandwich argument
yields $\sigma_{\mathrm{imp}}(K,\tau)\to v_L^\ast/\sqrt\tau$ as $K\to 0$.
\end{proof}

\subsection{Proof of proposition \ref{proposition:func in weighted Sobolev Space} }
Recall that for $m \in \mathbb{N}$, the weighted Sobolev space $\mathscr{H}_m$ with the weight function $\rho(x) = \frac{1}{\sqrt{2\pi\sigma^2}} e^{-\frac{x^2}{2\sigma^2}}$ consists of functions $f \in L_\rho^2$ whose weak derivatives up to order $m$ belong to $L_\rho^2$:
\begin{equation*}
\mathscr{H}_m := \left\{ f \in L_\rho^2 \mid \|f\|_m := \left( \sum_{\tau=0}^m \|f^{(\tau)}\|_{L_\rho^2}^2 \right)^{1/2} < \infty \right\},
\end{equation*}
where $L_\rho^2 = \left\{ f: \mathbb{R} \rightarrow \mathbb{R} \mid \|f\|_{L_\rho^2}^2 = \int_{\mathbb{R}} |f(x)|^2 \rho(x) \, \mathrm{d} x < \infty \right\}$.

By our construction, $F_{\mu_i}, F^{-1}_{\mu_i},F \in C^m$ are piece-wise polynomial functions, and the convolution expression $\int_{\mathbb{R}} \rho(y) \cdot F_{\mu_{i+1}}^{-1}\left(\int_{\mathbb{R}} F(w-y-x) \rho(x) \mathrm{d} x\right) \mathrm{d} y$ shows that the inner and outer integrand shares the same structure: $\int f(x)\cdot\rho(x) dx$. To show the integrands belongs to $\mathscr{H}_m$, we only need to focus on the generic function $f \in C^m(\mathbb{R})$ with at most polynomial growth because of the spline method. That is, for each $k=0,1, \ldots, m$, there exist constants $C_k, p_k>0$ such that:

$$
\left|f^{(k)}(x)\right| \leq C_k\left(1+|x|^{p_k}\right)
$$

for all $x \in \mathbb{R}$.

For each $k=0,1, \ldots, m$, we have:

$$
\int_{\mathbb{R}}\left|f^{(k)}(x)\right|^2 \rho(x) d x \leq C_k^2 \int_{\mathbb{R}}\left(1+|x|^{p_k}\right)^2 \rho(x) d x
$$

The integrand $\left(1+|x|^{p_k}\right)^2 \rho(x)$ behaves like $|x|^{2 p_k} e^{-\frac{x^2}{2 \sigma^2}}$ for large $|x|$. Since the Gaussian decay dominates any polynomial growth (that is, $\lim _{|x| \rightarrow \infty}|x|^p e^{-\frac{x^2}{2 \sigma^2}}=0$ for any power $p$ ), this integral is finite.

Therefore, $\int_{\mathbb{R}}\left|f^{(k)}(x)\right|^2 \rho(x) d x<\infty$ for all $k=0,1, \ldots, m$, and with only finite terms of summations, it means $f \in \mathscr{H}_m$.

For the Gauss-Hermite quadrature applied to functions with this level of regularity, classical results in numerical analysis (Brass \& Petras, 2011) establish that with $n$ quadrature points, the approximation error is bounded by:
\begin{align}
\left|\int_{\mathbb{R}} f(x)\rho(x)\,dx - \sum_{i=1}^n w_i f(x_i)\right| \leq C \cdot \|f^{(m)}\|_{L^2_\rho} \cdot n^{-m/2}
\end{align}
where $C$ is a constant depending only on $m$, and $\|f^{(m)}\|_{L^2_\rho} = \left(\int_{\mathbb{R}} |f^{(m)}(x)|^2 \rho(x) dx\right)^{1/2}$.

Thus, the convergence rate of Gauss-Hermite quadrature for the numerical evaluation of the integrals in equation \eqref{eq:double integrand} is $\mathcal{O}(n^{-m/2})$, where $m$ is the smoothness order of the functions involved.

Finally, by applying results from \citep{kazashi2023suboptimality} and \citep{mastroianni1994error}, we derive the convergence rate for Gauss-Hermite quadrature in the weighted Sobolev space \(\mathscr{H}_m\). Specifically, for a function \(f \in \mathscr{H}_m\), the Gauss-Hermite quadrature approximation \(Q_n^{\mathrm{GH}}(f)\) to the integral \(I(f) := \int_{\mathbb{R}} f(x) \rho(x) \, \mathrm{d}x\), where \(\rho(x)\) is the Gaussian weight function, satisfies the following error bound:
\[
\left| I(f) - Q_n^{\mathrm{GH}}(f) \right| \leq C n^{-m/2} \|f\|_m,
\]
where \(n\) is the number of quadrature points, \(m\) is the smoothness order of the function \(f\), \(\|f\|_m\) is the norm in the weighted Sobolev space, and \(C > 0\) is a constant independent of \(n\).

This result shows that the convergence rate of the Gauss-Hermite quadrature depends on both the number of quadrature points \(n\) and the smoothness \(m\) of the function \(f\). As \(n\) increases, the error decays at a rate proportional to \(n^{-m/2}\), with smoother functions (i.e., higher \(m\)) leading to faster convergence. The constant \(C\) is determined by the specific properties of the function and the quadrature scheme but remains independent of the number of quadrature points.

To be brief, we conclude that in Bass-LV construction, the rate of convergence for Gauss-Hermite quadrature under finite smoothness condition can achieve $\mathcal{O}\left(n^{-m / 2}\right)$. Now we complete the proof for this part.

\subsection{Proof of Proposition \ref{proposition:opt trap}}

Before deriving the optimality of the Trapezoidal Rule Scheme, we first present some properties of Hermite polynomials, which will be useful for our weighted function \(\rho(x) = \frac{1}{\sqrt{2\pi\sigma^2}} e^{-\frac{x^2}{2\sigma^2}}\). Recall that the probabilist's Hermite polynomials are defined as:
\[
H_{e_n}(x) = (-1)^n e^{\frac{x^2}{2}} \frac{d^n}{dx^n} e^{-\frac{x^2}{2}},
\]
and satisfy the orthogonality relation:
\[
\int_{-\infty}^{\infty} H_{e_m}(x) H_{e_n}(x) e^{-\frac{x^2}{2}} dx = \sqrt{2 \pi} n! \delta_{n m},
\]
where \(\delta_{n m}\) is the Kronecker delta. We now state the following proposition:

\begin{proposition}
The normalized Hermite polynomials with respect to \(\rho(x)\) are given by:
\[
H_{e_n}^{\sigma}(x) = \frac{(-1)^n \sigma^n}{\sqrt{n!}} e^{\frac{x^2}{2\sigma^2}} \frac{d^n}{dx^n} e^{-\frac{x^2}{2\sigma^2}}, \quad x \in \mathbb{R},
\]
and satisfy the recurrence relation:
\[
(H_{e_n}^{\sigma}(x))' = \frac{\sqrt{n}}{\sigma} H_{e_{n-1}}^\sigma(x), \quad \forall n \geq 1.
\]
\label{proposition:Hek normalized property}
\end{proposition}

\textbf{Proof:} First, we verify the normalization by calculating the \(L^2_\rho\) norm:
\[
\int_{-\infty}^{\infty} H_{e_n}^{\sigma}(x) H_{e_n}^{\sigma}(x) \rho(x) dx = 1.
\]
This follows directly from applying the orthogonality relation of the Hermite polynomials, and adjusting for the scaling factor \(\sigma\):

\begin{align*}
& \int_{-\infty}^{\infty} H_{e_n}^\sigma(x) H_{e_n}^\sigma(x) \frac{1}{\sqrt{2 \pi \sigma^2}} e^{-\frac{x^2}{2 \sigma^2}} d x \\
= & \frac{\sigma^{2 n}}{n!\sqrt{2 \pi}} \int_{-\infty}^{\infty}\left[e^{\frac{x^2}{2 \sigma^2}} \frac{d^n}{d x^n} e^{-\frac{x^2}{2 \sigma^2}}\right]^2 e^{-\frac{x^2}{2 \sigma^2}} \frac{1}{\sigma} d x \\
= & \frac{\sigma^{2 n}}{n!\sqrt{2 \pi}} \int_{-\infty}^{\infty}\left[e^{\frac{y^2}{2}} \frac{1}{\sigma^n} \frac{d^n}{d y^n} e^{-\frac{y^2}{2}}\right]^2 e^{-\frac{y^2}{2}} d y \\
= & \frac{1}{n!\sqrt{2 \pi}} \int_{-\infty}^{\infty} H_{e_n}(y) H_{e_n}(y) e^{-\frac{y^2}{2}} d y \\
= & \frac{1}{n!\sqrt{2 \pi}} \sqrt{2 \pi} n!=1    
\end{align*}

The recurrence relation is derived using the chain rule on the differentiated form of \(H_{e_n}^{\sigma}(x)\), applying the Leibniz rule for products of exponentials and polynomials:

\begin{align*}
\left(H_{e_n}^\sigma(x)\right)^{\prime} &= \frac{d}{dx}\left[\frac{(-1)^n \sigma^n}{\sqrt{n!}} e^{\frac{x^2}{2\sigma^2}} \frac{d^n}{dx^n} e^{-\frac{x^2}{2\sigma^2}}\right] \\
&= \frac{(-1)^n \sigma^n}{\sqrt{n!}}\left[\frac{x}{\sigma^2}e^{\frac{x^2}{2\sigma^2}} \frac{d^n}{dx^n} e^{-\frac{x^2}{2\sigma^2}} + e^{\frac{x^2}{2\sigma^2}} \frac{d^{n+1}}{dx^{n+1}} e^{-\frac{x^2}{2\sigma^2}}\right] \\
&= \frac{(-1)^n \sigma^n}{\sqrt{n!}}\left[\frac{x}{\sigma^2}e^{\frac{x^2}{2\sigma^2}} \frac{d^n}{dx^n} e^{-\frac{x^2}{2\sigma^2}} + e^{\frac{x^2}{2\sigma^2}} \left(-\frac{x}{\sigma^2} \frac{d^n}{dx^n}e^{-\frac{x^2}{2\sigma^2}} - \frac{n}{\sigma^2} \frac{d^{n-1}}{dx^{n-1}}e^{-\frac{x^2}{2\sigma^2}}\right)\right] \\
&= \frac{(-1)^n \sigma^n}{\sqrt{n!}} \left[- \frac{n}{\sigma^2}e^{\frac{x^2}{2\sigma^2}} \frac{d^{n-1}}{dx^{n-1}}e^{-\frac{x^2}{2\sigma^2}}\right] \\
&= \frac{(-1)^{n+1} n \sigma^{n-2}}{\sqrt{n!}} e^{\frac{x^2}{2\sigma^2}} \frac{d^{n-1}}{dx^{n-1}}e^{-\frac{x^2}{2\sigma^2}} \\
&= \frac{(-1)^{n+1} n \sigma^{n-2} \sqrt{(n-1)!}}{\sqrt{n} \cdot \sqrt{(n-1)!}} H_{e_{n-1}}^{\sigma}(x) \\
&= \frac{\sqrt{n}}{\sigma} H_{e_{n-1}}^{\sigma}(x)
\end{align*}

We will use this result to bound the norms of the function in the next part of the proof.

\textbf{Lemma 1 (Bounded Norms).} Assume the inner integrand of the Bass-LV construction has \(m\)-order smoothness. Let \(F_i(x) := F(w - x - y) \cdot \rho(x)\), where \(F\) is the distribution obtained in a previous iteration, \(w, y\) are constants, and \(\rho(x) = \frac{1}{\sqrt{2\pi\sigma^2}} e^{-\frac{x^2}{2\sigma^2}}\) is the Gaussian weight function. For the \(\tau\)-th order derivative of \(F_i(x)\), we have the following bounds:
\[
\|F_i^{(\tau)}(x)\|_{L^1(\mathbb{R})} < \infty, \quad \|F_i^{(m)}(x)\|_{L^2(\mathbb{R})} < \infty, \quad \sup_{x \in \mathbb{R}} \left| e^{(1-\varepsilon) \frac{x^2}{2\sigma^2}} F_i^{(\tau)}(x) \right| < \infty,
\]
where \(\varepsilon\) is a small positive constant such that \(\frac{1-\epsilon}{\sigma^2} \in (0, 1)\).

\textbf{Proof:}
Since our weighted function is heat kernel with zero drift, it aligns with the form of Hermite quadrature we constructed in proposition \ref{proposition:Hek normalized property}. We can write under chain rule that:
\begin{align*}
\|F_i^{(\tau)}\|_{L^1(\mathbb{R})}
&\leq \sum_{k=0}^\tau \binom{\tau}{k} \|F^{(\tau-k)}(w-x-y) \cdot \rho^{(k)}(x)\|_{L^1} \\
&= \sum_{k=0}^\tau \binom{\tau}{k} \left( \int_{\mathbb{R}} |F^{(\tau-k)}(w-x-y) \frac{(-1)^k}{\sigma^k} \sqrt{k!} H_{e_k}^\sigma(x)\frac{1}{\sqrt{2\pi\sigma^2}} e^{-\frac{x^2}{2\sigma^2}}| \, \mathrm{d}x \right) \\
&(\textbf{Applying H\"older's Inequality}) \\
&\leq \sum_{k=0}^\tau \binom{\tau}{k} \frac{\sqrt{k!}}{\sigma^k} \left( \int_{\mathbb{R}} |F^{(\tau-k)}(w-x-y)|^2 \rho(x) \, \mathrm{d}x \right)^{1/2} \left( \int_{\mathbb{R}} |H_{e_k}^\sigma(x)|^2 \rho(x) \, \mathrm{d}x \right)^{1/2}.
\end{align*}

Given the analysis in proposition \ref{proposition:func in weighted Sobolev Space}, we know that $F \in C^m(\mathbb{R})$ with at most polynomial growth. This means that $\left(\int_{\mathbb{R}}\left|F^{(\tau-k)}(w-x-y)\right|^2 \rho(x) \mathrm{~d} x\right)$ is bounded by some constant:
\begin{equation*}
\left(\int_{\mathbb{R}}\left|F^{(\tau-k)}(w-x-y)\right|^2 \rho(x) \mathrm{~d} x\right) < C^{(\tau-k)}_{F_w},
\end{equation*}
where $C^{(\tau-k)}_{F_w}$ is some positive constant. 

By proposition \ref{proposition:Hek normalized property}, we know that $\left(\int_{\mathbb{R}}\left|H_{e_k}^\sigma(x)\right|^2 \rho(x) \mathrm{d} x\right)=1<\infty$.
With these analyses, we can then obtain the following:
\begin{align}
\left\|F_i^{(\tau)}\right\|_{L^1(\mathbb{R})}\leq \sum_{k=0}^\tau \binom{\tau}{k} \frac{\sqrt{k!}}{\sigma^k} (C^{(\tau-k)}_{F_w})^{\frac{1}{2}} < \infty
\end{align}
Now we complete the first part of the proof.

For the $L^2$ norm, since $\int_{\mathbb{R}}\left|H_{e_k}^\sigma(x)\right|^2 \rho(x) \mathrm{d} x=1$, there must exist a supremum $C_H$ such that $\left|H_{e_k}^\sigma(x)\right|^2 \rho(x)<C_H,\forall x\in\mathbb{R}$. Therefore, we can write the following:
\begin{align*}
\left\|F_i^{(m)}\right\|_{L^2(\mathbb{R})}
&\leq \sum_{k=0}^m \binom{m}{k} \left(\int_{\mathbb{R}} \left|{F}^{(m-k)}(w-y-x)\frac{(-1)^k}{\sigma^k} \sqrt{k!}H_{e_k}^\sigma(x)\frac{1}{\sqrt{2\pi\sigma^2}}e^{-\frac{x^2}{2\sigma^2}}\right|^2 \, \mathrm{d}x\right)^{\frac{1}{2}}\\
&\leq \sum_{k=0}^m \binom{m}{k} \left(C_H\frac{k!}{\sigma^{2k}}\int_{\mathbb{R}} \left| {F}^{(m-k)}(w-y-x) \right|^2\rho(x)dx\right)^\frac{1}{2}\\
&< \sum_{k=0}^m \binom{m}{k} \left(C_H\frac{k!}{\sigma^{2k}}C_{F_w}^{(m-k)}\right)^\frac{1}{2}<\infty
\end{align*}

For the infinity norm, we denote \( e^{\frac{(1-\epsilon) x^2}{2 \sigma^2}} = \rho(x)^{\epsilon - 1} \). Note that \( H_{e_k}^\sigma(x) \) is a \( k \)-th order polynomial, i.e., the asymptotic behavior of \( H_{e_k}^\sigma(x) \) is \( O(x^k) \). We first establish that there exists a finite supremum \( C_H^{\epsilon_1} \) for the product \( F_H(x) := H_{e_k}^\sigma(x) \rho^{\epsilon_1}(x) \), where \( \epsilon_1 > 0 \) and \( x \in \mathbb{R} \).

Since \( \rho(x) = \frac{1}{\sqrt{2 \pi \sigma^2}} e^{-\frac{x^2}{2 \sigma^2}} \), for any \( \epsilon_1 > 0 \), \( \rho(x) \) decays exponentially as \( x \to \pm \infty \), while \( H_{e_k}^\sigma(x) \), as a polynomial, can increase or decrease at most polynomially in \( x \). Therefore, the product \( F_H(x) \) exhibits a finite bound as \( x \to \pm \infty \). 

Furthermore, for any sufficiently large but finite interval \( [a, b] \), \( F_H(x) \) is continuous by construction. By applying the extreme value theorem, we conclude that it attains finite upper and lower bounds on any such interval. Now consider \( G_W(x) := F^{(\tau)}(w - y - x) \rho^{\epsilon_2}(x) \), where \( \epsilon_2 > 0 \) and \( w, y \) are constants. Similar to the previous case, we examine the behavior of \( G_W(x) \) as \( x \to \pm \infty \).

From the analysis in proposition \ref{proposition:func in weighted Sobolev Space}, we know that \(F \in C^m(\mathbb{R})\) with at most polynomial growth. This means that for each \(k=0,1, \ldots, m\), there exist constants \(C_k, p_k>0\) such that \(|F^{(k)}(x)| \leq C_k(1+|x|^{p_k})\) for all \(x \in \mathbb{R}\). When combined with the exponential decay of \(\rho^{\epsilon_2}(x)\), the product \(G_W(x) := F^{(\tau)}(w - y - x) \rho^{\epsilon_2}(x)\) decays rapidly enough to ensure a finite supremum. This is because for any polynomial growth of \(F^{(\tau)}\), the exponential decay of \(\rho^{\epsilon_2}\) will dominate as \(|x| \to \infty\). By applying dominated convergence properties, we establish that \(G_W(x)\) has a finite supremum, denoted \(C^{\epsilon_2}_{F^{(\tau)}_w}\), for any given \(\epsilon_2 > 0\) and derivative order \(\tau\).

Thus, we conclude that for the supremum norm:
\[
\sup _{\substack{x \in \mathbb{R} \\ \tau \in\{0, \ldots, m-1\}}}\left|e^{(1-\varepsilon) \frac{x^2}{2 \sigma^2}} F_i^{(\tau)}(x)\right|
\]
we have:

\begin{align*}
&\left|\left|e^{(1-\varepsilon) \frac{x^2}{2\sigma^2}} F_i^{(\tau)}(x)\right|\right|_{L^\infty(\mathbb{R})}\\  
&\leq \sum_{k=0}^\tau \binom{\tau}{k} \left|\left|\rho(x)^{\epsilon-1} {F}^{(\tau-k)}(w-y-x)\rho^{(k)}(x)\right|\right|_{L^\infty(\mathbb{R})}\\
&=\sum_{k=0}^\tau \binom{\tau}{k} \left|\left|\rho(x)^{\epsilon-1} {F}^{(\tau-k)}(w-y-x)\rho(x)H_{e_k}^\sigma(x)\frac{(-1)^k\sqrt{k!}}{\sigma^k}\right|\right|_{L^\infty(\mathbb{R})}\\
&\leq\sum_{k=0}^\tau \binom{\tau}{k} \left|\left|\rho(x)^{\frac{\epsilon}{2}} {F}^{(\tau-k)}(w-y-x)\right|\right|_{L^\infty(\mathbb{R})}\left|\left|H_{e_k}^\sigma(x)\frac{(-1)^k\sqrt{k!}}{\sigma^k}\rho(x)^{\frac{\epsilon}{2}}\right|\right|_{L^\infty(\mathbb{R})}\\
&\leq\sum_{k=0}^\tau \binom{\tau}{k}C^{\frac{\epsilon}{2}}_H\frac{\sqrt{k!}}{\sigma^k}C^{\frac{\epsilon}{2}}_{F^{(\tau-k)}_w}
\end{align*}

By construction, we need choose $\epsilon \in (max\{0,1-\sigma^2\},1)$. In the last inequality, we obtain the supremum by setting $\epsilon_1=\epsilon_2=\frac{\epsilon}{2}$. Since $\tau \in [0,m-1]$ is finite, the finite summation of bounded values is finite, thus giving $\sup _{\substack{x \in \mathbb{R} \\ \tau \in\{0, \ldots, m-1\}}}\left|e^{(1-\varepsilon) \frac{x^2}{2 \sigma^2}} F_i^{(\tau)}(x)\right|<\infty$.

\textbf{Application of Theorem:} With the above lemma established, we now apply the results from \citep{kazashi2023suboptimality}, Proposition 4.2, to our numerical setup in the Bass-LV model. 

\textbf{Theorem (Kazashi 2023, Proposition 4.2):} 
Let $m \in \mathbb{N}$ represent the smoothness order of the function $g$. Assume that $g^{(\tau)}: \mathbb{R} \rightarrow \mathbb{R}$ is absolutely continuous on any compact interval for each derivative order $\tau = 0, \ldots, m-1$, and that the $m$-th derivative, $g^{(m)}$, belongs to the space $L^2(\mathbb{R})$ (the space of square-integrable functions). Additionally, $g$ must satisfy the following two conditions:

1. Local Regularity: The $m$-th order Sobolev norm of $g$, denoted as \( \|g\|_m^* \), is uniformly bounded over all compact intervals $I \subset \mathbb{R}$. This norm is defined as:
\[
\|g\|_m^* := \sup_{\substack{I \subset \mathbb{R} \\ |I| < \infty}} \|g\|_{m,I} := \sup_{\substack{I \subset \mathbb{R} \\ |I| < \infty}} \left( \sum_{\tau=0}^{m-1} \left( \int_I g^{(\tau)}(x) \, \mathrm{d}x \right)^2 + \int_I \left| g^{(m)}(x) \right|^2 \, \mathrm{d}x \right)^{1/2}.
\]
This norm ensures that $g$ and its derivatives up to order $m$ are well-behaved over compact intervals.

2. Decay at Infinity: The function $g$ must exhibit a controlled decay at infinity, expressed as:
\[
\|g\|_{m, \text{decay}} := \sup_{\substack{x \in \mathbb{R} \\ \tau \in \{0, \ldots, m-1\}}} \left| e^{(1-\varepsilon) \frac{x^2}{2}} g^{(\tau)}(x) \right| < \infty, \quad \text{for some } \varepsilon \in (0,1).
\]
This condition ensures that $g(x)$ and its derivatives decay rapidly enough as $x \to \pm \infty$, governed by the exponential decay factor $e^{(1-\varepsilon) \frac{x^2}{2}}$.

Given these assumptions, the error for the $n$-point Trapezoidal Rule Scheme $Q_{n,T}^*(g)$ with a cutoff interval $[-T, T]$ is bounded by:
\[
\left| \int_{\mathbb{R}} g(x) \, \mathrm{d} x - Q_{n, T}^*(g) \right| \leq C \left( \|g\|_m^* + \|g\|_{m, \text{decay}} \right) \frac{(\ln n)^{m/2 + 1/4}}{n^m},
\]
where $C$ is a constant independent of $n$ and $g$, but dependent on $m$ and $\varepsilon$. Here, the Trapezoidal Rule approximation $Q_{n, T}^*(g)$ is given by:
\[
Q_{n, T}^*(g) := \frac{2T}{n} \sum_{j=0}^{n-1} g\left( \xi_j^* \right),
\]
where $\xi_j^* := \frac{2T}{n} j - T$, and $T = \sqrt{\frac{2}{(1-\varepsilon)} m \ln n}$ is the cutoff interval.

\textbf{Adaptation for Bass-LV Model:} In the Bass-LV implementation, we modify the theorem to match the specifics of our integrand $g(x)$, which is given by:
\[
g(x) = F(w - y - x) \cdot \rho(x),
\]
where $\rho(x) = \frac{1}{\sqrt{2 \pi \sigma^2}} e^{-\frac{x^2}{2 \sigma^2}}$ is the weighted function representing the heat kernel. The prerequisite for $g(x)$ is satisfied in our proof above, the parameters $w$ and $y$ are constants, and $F(w - y - x)$ is the piecewise polynomial representation of the distribution function obtained in each iteration, which upon convergence of the algorithm, yields the final distribution $F_{W_{T_i}}$. 

We set \( T = Nh \), \( n = 2N + 1 \geq 2 \), and \( h = \frac{2T}{n} \), where $N$ is the number of quadrature points used. The term \( 1 - \varepsilon \) is adjusted to reflect the variance in the heat kernel, given by:
\[
1 - \varepsilon = \frac{1 - \epsilon}{\sigma^2}, \quad \epsilon \in (max\{1-\sigma^2,0\},1),
\]
where $\sigma^2 = T_{i+1} - T_i$ represents the time interval between maturities $T_i$ and $T_{i+1}$. 

Applying these modifications, we derive the optimal parameter settings for the Trapezoidal Rule Scheme for the inner integrand as follows:
\[
Nh = \sqrt{\frac{2(T_{i+1} - T_i)}{(1 - \epsilon)} m \ln(2N + 1)}, \quad h = \frac{\sqrt{\frac{2(T_{i+1} - T_i)}{(1 - \epsilon)} m \ln(2N + 1)}}{N}.
\]

\textbf{Outer Integrand:} The outer integrand, $F_{\mu_{i+1}}^{-1}$, is represented by a finite-order smoothness spline interpolation. Since the inputs to this function are also finite, similar results for optimality can be applied to the outer integrand, by imposing its own smoothness order and choosing $\epsilon$ independently within the interval \( \epsilon \in (\max \{ 1 - \sigma^2, 0 \}, 1) \).

\textbf{Convergence Rate:} From the results above, we conclude that the convergence rate of the Trapezoidal Rule Scheme for a single integrand in the Bass-LV implementation is:
\[
\mathcal{O}\left( \frac{(\ln n)^{m / 2 + 1 / 4}}{n^m} \right),
\]
where $m$ is the smoothness order of the integrand, and $n$ represents the number of points counted in the Trapezoidal Rule.

\bibliographystyle{unsrtnat}
\bibliography{references}

\section*{Disclaimer}
This paper was prepared for informational purposes with contributions from the Quantitative Trading \& Research team of JPMorgan Chase \& Co. This paper is not a product of the Research Department of JPMorgan Chase \& Co. or its affiliates. Neither JPMorgan Chase \& Co. nor any of its affiliates makes any explicit or implied representation or warranty and none of them accept any liability in connection with this paper, including, without limitation, with respect to the completeness, accuracy, or reliability of the information contained herein and the potential legal, compliance, tax, or accounting effects thereof. This document is not intended as investment research or investment advice, or as a recommendation, offer, or solicitation for the purchase or sale of any security, financial instrument, financial product or service, or to be used in any way for evaluating the merits of participating in any transaction.

\end{document}